\pdfoutput=1

\documentclass[a4paper,11pt]{article}

\usepackage{jheppub} 
\usepackage{xcolor,epsfig,amsmath,amssymb,subfigure,placeins}
\usepackage{graphicx,amsmath,bbm,bm,array}

\usepackage[normalem]{ulem}  % \sout{old text} for strikeout

\renewcommand\sout{\bgroup \color{red} \ULdepth=-.5ex \ULset}

\title{
	Spectral functions of heavy quarkonia \\ 
	in a bulk-viscous quark gluon plasma 
}
\author[a]{Lata Thakur}
\emailAdd{lata.thakur@apctp.org}

\author[a,b,c]{and Yuji Hirono}
\emailAdd{yuji.hirono@gmail.com}

\affiliation[a]{
	Asia Pacific Center for Theoretical Physics, \\
	Pohang, Gyeongbuk 37673, Republic of Korea
}

\affiliation[b]{
	Department of Physics, POSTECH, \\
	Pohang, Gyeongbuk 37673, Republic of Korea
}
\affiliation[c]{RIKEN iTHEMS, RIKEN, Wako 351-0198, Japan}

\abstract{
	We study the properties of quarkonia 
	inside a bulk-viscous quark gluon plasma. 
	The non-equilibrium nature of the medium 
	is encoded in the deformed 
	distribution functions of thermal quarks and gluons, 
	with which we compute the 
	dielectric permittivity within the hard thermal loop approximation at one-loop. 
	The modified dielectric permittivity is used to calculate the 
	in-medium heavy quark potential, 
	and using the potential 
	we compute spectral functions, which reflect the physical 
	properties of heavy quarkonia. 
	We discuss how the bulk viscous effect influences 
	quantities such as binding energies and thermal widths. 
	Based on those properties, 
	we discuss the implications of the bulk viscous effect on the physical observables such as $ \psi' $ to $ J/\psi $ ratio and the nuclear modification factor, $R_{AA}$. 
	In particular, 
	we argue that the nuclear modification factors of 
	excited and ground states 
	show different sensitivities to 
	the bulk viscous nature of a plasma, 
	which is potentially useful for the critical point search. 
}


\begin{document}
	\maketitle

	\section{Introduction}

	The suppression of heavy quarkonia \cite{Brambilla:2010cs,Rothkopf:2019ipj} is 
	one of the first proposed signals 
	for the formation of a quark gluon plasma (QGP) 
	in heavy ion collision experiments. 
	The properties of heavy quarkonia in the vacuum 
	are well-described by non-relativistic potential models \cite{Lucha:1991vn} 
	with a Coulombic behaviour at short distances and a non-perturbative linear rise at large distances~\cite{Eichten:1974af,Eichten:1978tg,Eichten:1979ms,Buchmuller:1980su,Koma:2006si,Brambilla:2004jw}. 
	At finite temperatures, inter-quark forces are modified 
	because of the medium effects. 
	It has been argued that above the crossover temperature, $ T_c $, the screening is strong enough 
	to dissociate $J/\psi$ state~\cite{Matsui:1986dk}. 
	The melting of quarkonia states is predicted to occur
	sequentially~\cite{Karsch:2005nk}. 
	The heavy ion collision experiments
	at RHIC and LHC have also observed relative suppressions of higher states 
	for bottomonia~\cite{CMS:2011all,CMS:2012gvv} and charmonia~\cite{ALICE:2012jsl}.

	The non-equilibrium aspects of a QGP 
	are getting an increasing attention these days, 
	and they may have some impact 
	on the properties of heavy quarkonia. 
	For example, the effect of momentum-space anisotropies 
	because of viscous effects 
	~\cite{Dumitru:2009ni,Dumitru:2009fy,Philipsen:2009wg,Dumitru:2010id,Margotta:2011ta,Thakur:2013nia,Thakur:2012eb,Strickland:2011aa,Strickland:2013uga,Strickland:2014pga,Agotiya:2016bqr} 
	or 
	external magnetic fields~\cite{Bonati:2015dka,Bonati:2016kxj,Marasinghe:2011bt,Alford:2013jva,Singh:2017nfa,Hasan:2017fmf,CS:2018jql} 
	have been discussed. 
	The relative motion of a fluid and heavy quarkonia 
	also has an impact on their  properties~\cite{Escobedo:2011ie,Escobedo:2013tca,Thakur:2016cki,Avramis:2006em,Liu:2012zw,Ali-Akbari:2014vpa,Patra:2015qoa,Lafferty:2019jpr}. 
	Among such effects, 
	the bulk viscous effect is 
	particularly of interest 
	in relation to the beam energy scan program~\cite{Bzdak:2019pkr}, 
	because it is expected to be enhanced 
	near a critical point~\cite{Kharzeev:2007wb,Karsch:2007jc,Moore:2008ws,Noronha-Hostler:2008kkf,Ryu:2015vwa,Monnai:2016kud}. 
	The influence of a bulk viscosity 
	on the time evolution of a plasma has been discussed, for example, in refs.~\cite{Monnai:2009ad,Denicol:2009am,Song:2009rh,Dusling:2011fd,Bozek:2012qs,Noronha-Hostler:2013gga,Noronha-Hostler:2014dqa,Bozek:2009dw}.

	In this work, we study the effect of bulk viscosity 
	on the properties of heavy quarkonia 
	and its implications on experimental observables
	such as the $\psi^{\prime}/J/\psi$ ratio and 
	the nuclear modification factor $R_{AA}$. 
	We incorporate the bulk viscous correction 
	through the modification of quasiparticle distribution functions in the momentum space. 
	This in turn modifies the dielectric permittivity, 
	which can be used \cite{Thakur:2020ifi} to obtain an in-medium potential
	for a bulk-viscous medium.\footnote{ 
		The modified complex potential contains both the perturbative Coulombic as well as
		the non-perturbative string-like terms.
		%,
		There are different ways for incorporating 
		non-perturbative terms in the potential~\cite{Thakur:2012eb,Thakur:2013nia,Agotiya:2008ie,Srivastava:2018vxp,Guo:2019bwa,Burnier:2015nsa,Lafferty:2019jpr}. 
		Non-perturbative contributions are 
		also observed in lattice QCD~\cite{Cheng:2008bs,Maezawa:2007fc,Andreev:2006eh}. 
	} 
	Using the modified heavy quark complex potential, 
	we compute the spectral functions \cite{Burnier:2007qm} 
	of quarkonium states in a bulk viscous medium. 
	From those spectral functions, 
	we can read off physical properties such as binding energies, resonance masses and decay widths of quarkonium states.
	We also compute $ \psi^{\prime}/J/\psi $ ratio and 
	nuclear modification factor $R_{AA}$, 
	which are measurable in heavy ion collision experiments.

	Let us state the difference of the current work 
	with our former one \cite{Thakur:2020ifi} 
	in which the effect of bulk viscosity on 
	heavy quarkonia is studied. 
	In ref.~\cite{Thakur:2020ifi}, 
	the dielectric permittivity is computed in the 
	Hard Thermal Loop (HTL) approximation using the thermal  distribution of massless quarks and gluons. 
	Here, we employ a quasiparticle model, 
	in which quasiparticles have 
	masses that depend on the temperature and chemical potential. 
	This model has a better description of thermodynamic quantities near the crossover temperature. 
	In this work, we compute the spectral functions 
	with a complex potential in medium 
	with a method developed in ref.~\cite{Burnier:2007qm}, 
	while in ref.~\cite{Thakur:2020ifi} 
	we solved the 
	Schr\"odinger equation using the real part of the potential
	and used the solved wave function to compute decay widths, 
	which may not be reliable when the imaginary part of the potential is large. 
	Here, we will also discuss implications on 
	experimental observables, 
	$ \psi^{\prime}/J/\psi $ ratio and $R_{AA}$.

	The rest of this article is structured as follows. 
	In section~\ref{sec:permittivity},
	we give the derivation of Debye masses and color dielectric permittivity for a bulk viscous medium.
	In section~\ref{sec:quarkonia}, 
	we discuss the spectral functions of quarkonia 
	in a bulk viscous medium. 
	In section~\ref{sec:implications}, 
	we discuss how the physical observables, 
	$ \psi^{\prime}/J/\psi $ ratio 
	and nuclear modification factor $R_{AA}$,
	are affected by the bulk viscous corrections. 
	Section \ref{sec:summary} is devoted to a summary.

	\section{Color dielectric permittivity of a bulk viscous QCD plasma}
	\label{sec:permittivity}
	
	In this section, 
	we give the computation of the 
	color dielectric permittivity of a bulk viscous medium, 
	which will be used later to get an in-medium potential. 
	First, we introduce the quasiparticle model, 
	and then we discuss how we compute the Debye masses and dielectric permittivity from the modified gluon self energies and propagators 
	following ref.~\cite{Du:2016wdx}.

	\subsection{Quasiparticle model}\label{sec:QPM}
	
	The quasiparticle model has been used 
	in the study of the thermal properties of a QGP. 
	It reproduces the behaviours of thermodynamic quantities 
	near the cross over temperature $T_{c}$, 
	where the perturbative QCD is not reliable. 
	In this model, the system of 
	interacting massless partons (quarks and gluons) is  effectively
	described as an ideal gas of massive non-interacting
	quasiparticles. 
	The quasiparticle mass, $m$, is given 
	a dependence on the temperature and chemical potential, 
	which arises due to the interactions of partons with the surrounding medium. 
	Such a functional dependence of quasiparticle mass turns out to reproduce the lattice data reasonably well 
	at finite temperatures. 
	This model was introduced by Goloviznin and Satz \cite{Goloviznin:1992ws} and further studies by 
	Peshier {\it et al.} \cite{Peshier:1995ty,Peshier:2002ww} 
	to explain the equation of states
	of a QGP obtained from lattice gauge simulation of QCD at 
	finite temperatures. 
	Simultaneously,
	different authors in refs.~\cite{Plumari:2011mk,Bluhm:2004xn,Bluhm:2007cp,Braaten:1991gm,Thakur:2019bnf,Srivastava:2015via,Thakur:2017hfc,Bannur:2006ww,Peshier:1999ww,Chandra:2009jjo} discussed the high-temperature lattice data
	by using a suitable quasiparticle description for QGP.

	In this work, we give the temperature ($T$), 
	and chemical potential ($\mu$) dependence 
	of the quasiparticle mass by~\cite{Peshier:1995ty,Peshier:2002ww,Peshier:1999ww,Bannur:2006js} 
	\begin{equation}
		m^{2}(T,\mu)=\frac{G^2(T)T^2 N_{c}}{9}+\frac{G^2(T)T^2 N_{f}}{18}\left(1+\frac{3 \mu^2}{\pi^2T^{2}}\right),
		\label{mth}
	\end{equation}
	where $ G^2(T) $ is given by~\cite{Peshier:1995ty} 
	\begin{equation}
		G^2(T)=\frac{48 \pi^2}{(11 N_c-2N_f)\ln\left(\lambda \frac{T}{T_c}+\frac{T_s}{T_c}\right)^{2}}.
	\end{equation}
	The parameters are chosen as 
	$ N_c=3 $, $ N_f=3 $, $ \lambda=4.17 $, 
	$ T_s/T_c=-2.96 $, and 
	$ T_c=160  $~MeV.

	\subsection{Debye masses and color dielectric permittivity} \label{sec:debyemass}
	
	In this subsection, we compute the Debye masses and dielectric permittivity of a bulk viscous plasma. 
	We introduce the bulk viscous correction
	by deforming the distribution functions 
	of thermal quasiparticles. 
	We parametrize the deformed distribution function 
	in the following way \cite{Du:2016wdx}\footnote{
		This form is motivated 
		by the following deformation of the distribution function 
		\cite{Nopoush:2014pfa, Du:2016wdx} 
		\begin{equation*}
			f(k)=f^{\rm id}\left(\frac{1}{T}\sqrt{k^{2}+m^{2}(1+\Phi )}\right). 
		\end{equation*}
	}
	\begin{equation}
		\begin{split}
			f(k) 
			&=
			f^{\rm id}(\tilde{k}) +\delta_{\rm bulk}f(\tilde{k}) \\
			&=
			f^{\rm id}(\tilde{k})+\frac{m^{2}\Phi}{2T\sqrt{k^2+m^2}}f^{\rm id}(\tilde{k})(1\pm f^{\rm id}(\tilde{k})) , 
		\end{split}
		\label{fk}
	\end{equation}
	where $ f^{\rm id} $ is the isotropic reference distribution (without non-equilibrium corrections)
	and  $\tilde{k}\equiv \frac{1}{T}\sqrt{k^{2}+m^{2}}  $. Here $+(-)$ sign is for Bose (Fermi) distribution, 
	$ m $ is the quasiparticle mass, 
	whose temperature- and chemical potential-dependence 
	is given by eq.~(\ref{mth}). 
	Here, the parameter $\Phi$ quantifies 
	the strength of the bulk-viscous correction.
	We here compute the propagators in the presence of 
	the correction \eqref{fk}. 
	It will turn out that 
	the effect of the bulk viscous correction is 
	to shift the Debye masses 
	for the retarded, advanced, and symmetric self energies.

	Let us first look at the equilibrium contribution to the retarded self energy. 
	In the HTL approximation, 
	the one-loop quarks contribution to the retarded self energy, $\Pi_R(P)$ in the presence of mass for $ \Phi=0 $ is written as~\cite{Mrowczynski:2000ed} 
	\begin{equation}
		\Pi^{\rm id,q}_{R}(P)=N_f\frac{g^2 }{\pi^2} \int dk k \left(f^{\rm id}_{+}(\tilde{k}) +f^{\rm id}_{-}(\tilde{k})\right) 
		\left(\frac{p^{0}}{2p}\ln\frac{p^{0}+p+i\epsilon}{p^{0}-p+i\epsilon}-1\right), 
	\end{equation}
	where $ f^{\rm id}_{\pm}(\tilde{k}) $ are the distribution functions for quarks/antiquarks in the thermal equilibrium, which is given by 
	\begin{equation}
		f^{\rm id}_{\pm}(\tilde{k})
		=\frac{1}{e^{\left( \sqrt{k^{2}+m^{2}}\mp \mu\right) /T}+ 1}\,. 
	\end{equation}
	The integral can be performed to give 
	\begin{equation}
		\Pi^{\rm id,q}_{R}(P)= N_f\frac{g^2 T^2}{6}\left(1+\frac{3 \mu^2}{\pi^2T^{2}}\right)f_{q}(\tilde{m},\tilde{\mu})\left(\frac{p^{0}}{2p}\ln\frac{p^{0}+p+i\epsilon}{p^{0}-p+i\epsilon}-1\right),  
	\end{equation}
	where $ \tilde{m}=m/T$, $  \tilde{\mu}=\mu/T$ and $ f_{q}(\tilde{m},\tilde{\mu}) $ is
	%where
	\begin{equation}
		f_{q}(\tilde{m},\tilde{\mu})=2\left[1-
		\frac{3\tilde{m}\tilde{\mu}-3\tilde{m} \ln [(1+e^{\tilde{m}+\tilde{\mu}})(1+e^{\tilde{\mu}-\tilde{m}})]-3[{\rm Li}_{2}(-e^{\tilde{m}-\tilde{\mu}})]+{\rm Li}_{2}(-e^{\tilde{m}+\tilde{\mu}})}{\pi^2+3\tilde{\mu}^2}\right], 
	\end{equation}
	where ${\rm Li}_n(x)$ denotes the polylogarithm function.
	Similarly, one can compute the gluon contribution to the retarded self energy 
	using the gluon distribution function in the presence of quasiparticle mass (defined in eq.~\eqref{mth}) as
	\begin{equation}
		f_{g}^{\rm id}(\tilde{k})
		=\frac{1}{e^{ \frac{1}T \sqrt{k^{2}+m^{2}} }- 1}\,, 
	\end{equation}
	as 
	\begin{equation}
		\Pi^{\rm id,g}_{R}(P)= 2 N_c\frac{g^2 T^2}{6}f_{g}(\tilde{m})\left(\frac{p^{0}}{2p}\ln\frac{p^{0}+p+i\epsilon}{p^{0}-p+i\epsilon}-1\right), 
	\end{equation}
	where $ f_g(\tilde{m})$ is given by 
	\begin{eqnarray}
		f_{g}(\tilde{m})
		= 
		\frac{1}{\pi^2} 
		\left( 3\tilde{m}^2+2\pi^2-6\tilde{m} \ln [e^{\tilde{m}}-1]-6{\rm Li}_{2}(e^{\tilde{m}}) \right) . 
	\end{eqnarray}
	Therefore, the total retarded self energy 
	at the thermal equilibrium is~\cite{Du:2016wdx} 
	\begin{eqnarray}
		\Pi^{\rm id}_{R}(P)&=&\Pi^{\rm id,q}_{R}(P)+\Pi^{\rm id,g}_{R}(P)\nonumber\\
		&=& m^2_{D,R}\left(\frac{p^{0}}{2p}\ln\frac{p^{0}+p+i\epsilon}{p^{0}-p+i\epsilon}-1\right),
	\end{eqnarray}
	where $ m_{D,R} $ is 
	the retarded Debye mass at $ \Phi=0 $, 
	\begin{equation}
		m^2_{D,R}=\frac{g^2T^2}{6}\left( N_f f_{q}(\tilde{m},\tilde{\mu})~\left(1+\frac{3\tilde{\mu}^2}{\pi^2}\right) +2N_cf_{g}(\tilde{m})\right) . 
		\label{mDReq}
	\end{equation}
	In the massless case, above equation reduces to the familiar expression, 
	\begin{equation}
		m^2_{D,R}=m^2_{D}=\frac{g^2T^2}{6}\left( N_f ~\left(1+\frac{3\tilde{\mu}^2}{\pi^2}\right) +2N_c\right) . 
		\label{mDeq}
	\end{equation}
	%%%%%%%%%%%%%%%%%%%%%%%%%%
	
	Let us now look at the bulk viscous correction. 
	The quark contribution to the 
	retarded self energy in the presence of bulk correction reads 
	\begin{eqnarray}
		\delta_{\rm bulk}\Pi^{(q)}_{R}(P)&=& N_f\frac{g^2 }{\pi^2}\frac{m^2\Phi}{2 T} \int dk \frac{k}{\sqrt{k^2+m^2}} \left(f^{\rm id}_{+}(\tilde{k})\left(1-f^{\rm id}_{+}(\tilde{k})\right)
		+f^{\rm id}_{-}(\tilde{k})\left(1-f^{\rm id}_{-}(\tilde{k})\right)\right) \nonumber\\ &\times&\left(\frac{p^{0}}{2p}\ln\frac{p^{0}+p+i\epsilon}{p^{0}-p+i\epsilon}-1\right)\nonumber\\
		&=& N_f\frac{g^2T^2 }{6}\frac{\tilde{m}^2}{\pi^2}\Phi \left( \frac{3}{1+e^{\tilde{m}- \tilde{\mu}}}+\frac{3}{1+e^{\tilde{m}+\tilde{\mu}} }\right) \left(\frac{p^{0}}{2p}\ln\frac{p^{0}+p+i\epsilon}{p^{0}-p+i\epsilon}-1\right), \nonumber\\
	\end{eqnarray}
	and the gluon contribution is 
	\begin{equation}
		\delta_{\rm bulk}\Pi^{g}_{R}(P)
		= 2 N_c\frac{g^2 T^2}{6} \frac{\tilde{m}^2}{\pi^2 }\Phi \left( \frac{3}{e^{\tilde{m}}- 1}   \right)  \left(\frac{p^{0}}{2p}\ln\frac{p^{0}+p+i\epsilon}{p^{0}-p+i\epsilon}-1\right).
	\end{equation}
	The bulk viscous correction to the retarded self energy can be written as 
	\begin{equation}
		\delta_{\rm bulk}\Pi_{R}(P)=\delta m^2_{D,R}\left(\frac{p^{0}}{2p}\ln\frac{p^{0}+p+i\epsilon}{p^{0}-p+i\epsilon}-1\right),
	\end{equation}
	where $ \delta m^2_{D,R} $ is given by 
	\begin{equation}
		\delta m^2_{D,R}=\frac{g^2 T^2}{6}\Phi\left[  N_f\frac{\tilde{m}^2}{\pi^2} \left( \frac{3}{1+e^{\tilde{m}- \tilde{\mu}}}+\frac{3}{1+e^{\tilde{m}+\tilde{\mu}} }\right) +2N_c\frac{3\tilde{m}^2}{\pi^2 } \left( \frac{1}{e^{\tilde{m}}- 1}  \right)  \right] . 
	\end{equation}
	Hence, the total retarded self energy in the presence of bulk viscous corrections can be written as 
	\begin{eqnarray}
		\Pi_{R}(P)&=&	\Pi^{\rm id}_{R}(P)+\delta_{\rm bulk}\Pi_{R}(P)\nonumber\\
		&=&\widetilde{m}^2_{D,R}\left(\frac{p^{0}}{2p}\ln\frac{p^{0}+p+i\epsilon}{p^{0}-p+i\epsilon}-1\right),
		\label{PiR}
	\end{eqnarray}
	where $\widetilde{m}^2_{D,R}=m^2_{D,R}+\delta m^2_{D,R}$. 
	The retarded propagator can be obtained from the retarded self energy. 
	Using eq.~(\ref{PiR}), 
	we can calculate the temporal component 
	of the resummed retarded propagator in the static limit, $p_0  \to 0$, as  
	\begin{equation}
		\bar D_{R} (P)=
		\frac{1}{p^2+\widetilde m^2_{D, R}} . 
		\label{DRphi}
	\end{equation}
	The advanced propagator can be obtained by the complex conjugate
	of the retarded propagator.
	
	In a similar manner, 
	we can compute the symmetric self energy $\Pi_S(P)$ and the symmetric propagator $D_{S} (P)  $.
	The quark contribution to 
	the symmetric self energy 
	for $ \Phi=0 $~\cite{Mrowczynski:2000ed,Du:2016wdx} is 
	\begin{eqnarray}
		\Pi^{\rm id,q}_{S}(P)
		&=&4 i N_fg^2 \pi^2\int \frac{k^2 dk}{(2\pi)^3}\left( f_+^{\rm id}(\tilde{k})(f_+^{\rm id}(\tilde{k})-1)+ f_-^{\rm id}(\tilde{k})(f_-^{\rm id}(\tilde{k})-1)  \right)\frac{2}{p}\Theta(p^2-p^2_{0})\nonumber\\
		&=&-2\pi i N_f\frac{g^2T^2 }{6}\frac{T}{p}\frac{3}{\pi^2T^2}\int \frac{2k^2+m^2 }{\sqrt{k^2+m^2}}dk\left( f_{+}^{\rm id}(\tilde{k})+f_{-}^{\rm id}(\tilde{k}) \right)\Theta(p^2-p^2_{0}) \nonumber \\
		&=& 
		-2\pi i N_f\frac{g^2T^2 }{6}\frac{T}{p}~\frac{6m^2}{\pi^2T^2}\Theta(p^2-p^2_{0})\sum_{n=1}^{\infty}(-1)^{n+1}K_{2}(\tilde{m}n)\cosh(\tilde{\mu}n),
	\end{eqnarray}
	where $\Theta(x)$ is the step function,
	and ${\rm K}_n(x)$ denotes the modified Bessel function of the second kind.
	The gluon contribution to the symmetric self energy gives
	\begin{equation}
		\Pi^{\rm id,g}_{S}(P)=-2\pi i2 N_c\frac{g^2T^2 }{6}\frac{T}{p}~\frac{3m^2}{\pi^2T^2}\Theta(p^2-p^2_{0})\sum_{n=1}^{\infty}K_{2}(\tilde{m}n).
	\end{equation}
	The total symmetric self energy at $\Phi=0$ is 
	\begin{eqnarray}
		\Pi^{\rm id}_{S}(P)&=&\Pi^{\rm id,q}_{S}(P)+\Pi^{\rm id,g}_{S}(P)\nonumber\\
		&=& -2\pi i m^2_{D,S} \frac{T}{p}~\Theta(p^2-p^2_{0}),
	\end{eqnarray}
	where $ m_{D,S} $ is the symmetric Debye mass, 
	\begin{equation}
		m^2_{D,S}=\frac{g^2 T^2}{6}\left[N_f\frac{6\tilde{m}^2}{\pi^2}\sum_{n=1}^{\infty}(-1)^{n+1}\cosh(\tilde{\mu}n) K_{2}(\tilde{m}n)
		+2N_c  \frac{3\tilde{m}^2}{2\pi^2}\sum_{n=1}^{\infty}K_{2}(\tilde{m}n) \right].
	\end{equation}
	In the absence of a quasiparticle mass ($m=0$) 
	and a bulk viscous correction ($\Phi=0$), 
	the symmetric Debye mass reproduces the familiar expression, 
	\begin{equation}
		m^2_{D,S} = m^2_{D}=\frac{g^2T^2}{6}\left( N_f ~(1+\frac{3\tilde{\mu}^2}{\pi^2}) +2N_c\right).
		\label{mDeq1}
	\end{equation}
	To the first order in $ \Phi $, 
	the bulk viscous correction to the quark contribution of symmetric self energy reads 
	\begin{eqnarray}
		\delta_{\rm bulk}\Pi^{q}_{S}(P)
		&=&2\pi i N_f\frac{g^2T^2}{6}\frac{T}{p}~ \frac{3m^2\Phi}{T^42\pi^2 } \int \frac{k^2 dk}{\sqrt{k^2+m^2}}\left[\left( f_+^{\rm id}(\tilde{k})(f_+^{\rm id}(\tilde{k})-1)  \right)(1-2f_+^{\rm id}(\tilde{k}))\right.\nonumber\\
		&&\left.+\left( f_-^{\rm id}(f_-^{\rm id}(\tilde{k})-1)  \right)(1-2f_-^{\rm id}(\tilde{k}))\right]\Theta(p^2-p^2_{0})\nonumber\\
		&=&-2\pi i N_f\frac{g^2T^2 }{6}\frac{T}{p}~\frac{3m^3\Phi}{\pi^2T^3}\Theta(p^2-p^2_{0})\sum_{n=1}^{\infty}n(-1)^{n+1}K_{1}(\tilde{m}n)\cosh(\tilde{\mu}n),\,\,\,\,\,
	\end{eqnarray}
%%%%%%%%%%%%%%%%%%%%%
%
	and the correction to the gluon contribution is 
	\begin{equation}
		\delta_{\rm bulk}\Pi^{g}_{S}(P)=-2\pi i2 N_c\frac{g^2T^2 }{6}\frac{T}{p}~\frac{3m^3\Phi}{2\pi^2T^3}\Theta(p^2-p^2_{0})\sum_{n=1}^{\infty}K_{1}(\tilde{m}n).
	\end{equation}
	The total bulk viscous correction to the symmetric self energy is
	\begin{equation}
		\delta_{\rm bulk}\Pi_{S}(P)
		=
		\delta_{\rm bulk}\Pi^{q}_{S}(P)+\delta_{\rm bulk}\Pi^{g}_{S}(P)
		=-2\pi i~\delta m^2_{D,S}~\frac{T}{p}~\Theta(p^2-p^2_{0}), 
	\end{equation}
	where $\delta m^2_{D,S}$ is defined by  
	\begin{equation}
		\delta m^2_{D,S}=\frac{g^2 T^2}{6} \Phi \left[N_f\frac{6\tilde{m}^2}{\pi^2}\sum_{n=1}^{\infty}(-1)^{n+1}\cosh(\tilde{\mu}n) \frac{\tilde{m}}{2}nK_{1}(\tilde{m}n) 
		+2N_c  \frac{3\tilde{m}^2}{2\pi^2}\sum_{n=1}^{\infty} \frac{\tilde{m}}{2}nK_{1}(\tilde{m}n)\right].
	\end{equation}
	Thus, the effect of the bulk viscous correction enters as a shift of the Debye mass, and 
	the total symmetric self energy can be written as  
	\begin{eqnarray}
		\Pi_{S}(P)&=&\Pi^{\rm id}_{S}(P)+\delta_{\rm bulk}\Pi_{S}(P)\nonumber\\
		&=&-2\pi i~\widetilde{m}^2_{D,S}~\frac{T}{p}~\Theta(p^2-p^2_{0}),
		\label{PiS}
	\end{eqnarray}
	where $\widetilde{m}_{D,S}$ is defined by 
	$\widetilde{m}^2_{D,S}=m^2_{D,S}+\delta m^2_{D,S}$. 
	In the computation of the Debye masses, 
	the coupling constant 
	$ g^2= 4\pi \alpha_{s} $
	is evaluated 
	with the one-loop expression, 
	\begin{equation}
		\alpha_{s}
		=\frac{12 \pi}{(11N_{c}-2 N_{f})\ln\left(\frac{M^{2}}{\Lambda^{2}}\right)}, 
	\end{equation}
	and we take $ N_c=3 $, $ N_f=3 $, $\Lambda=176 $ MeV and $ M\approx 3.7 T$~\cite{Peshier:1995ty}.

	We can calculate the resummed symmetric propagator by using the relation
	$\bar D_S = \bar D_R  \Pi_S \bar D_A$.
	In the $p_0 \to 0$ limit, we have
	\begin{equation}
		\bar D_S
		=-\frac{2\pi i T  
			\widetilde{m}_{D,S}^2
		}{p(p^{2}+
			\widetilde{m}_{D,R}^2
			)^{2}}  
		. 
		\label{DSphi}
	\end{equation}
	%%%%%%%%%%%%%%
		\begin{figure}[tb]
		\subfigure{
			\hspace{-8mm}
			\includegraphics[width=7.8cm]{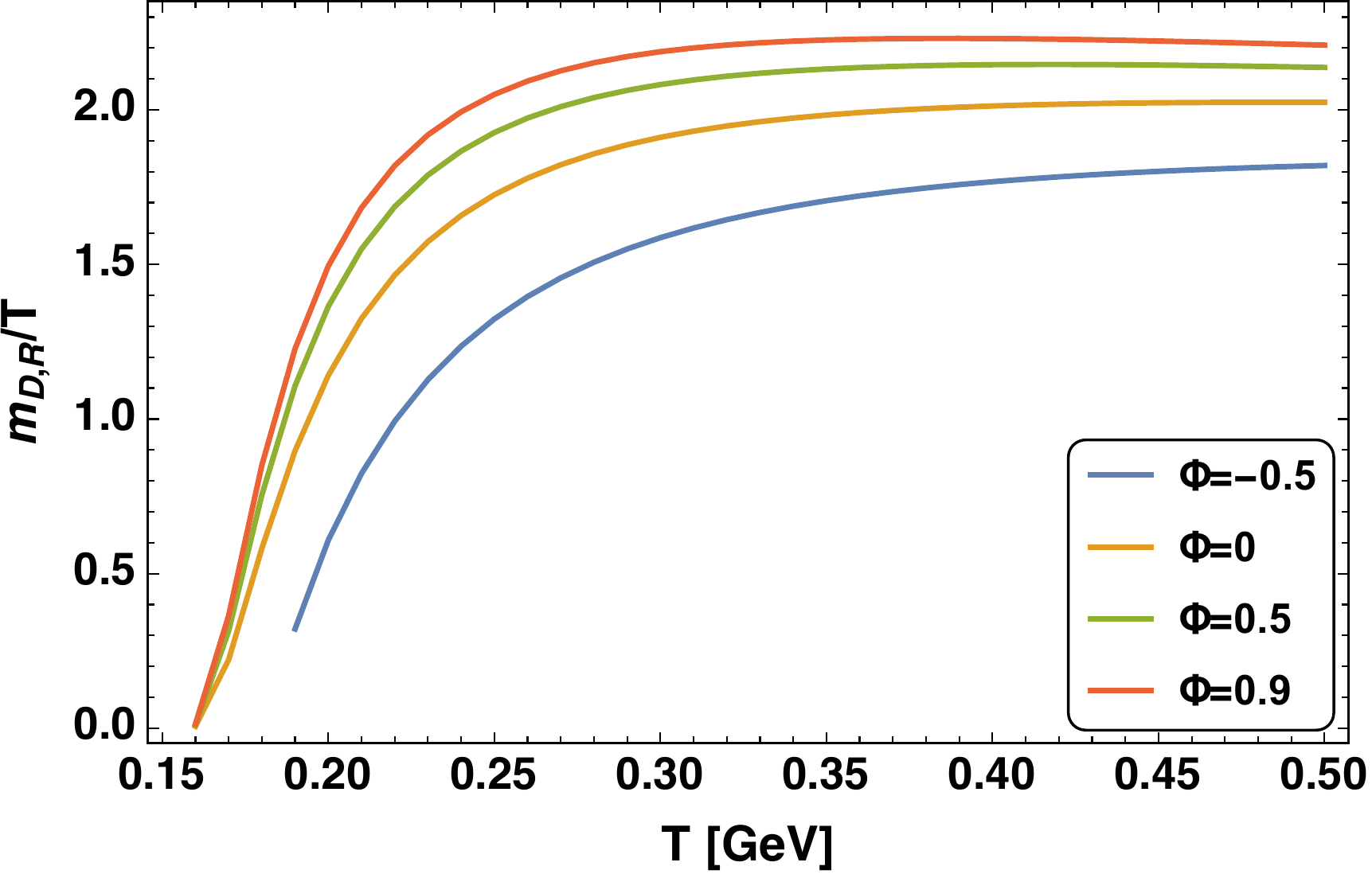}
		}
		\subfigure{
			\includegraphics[width=7.8cm]{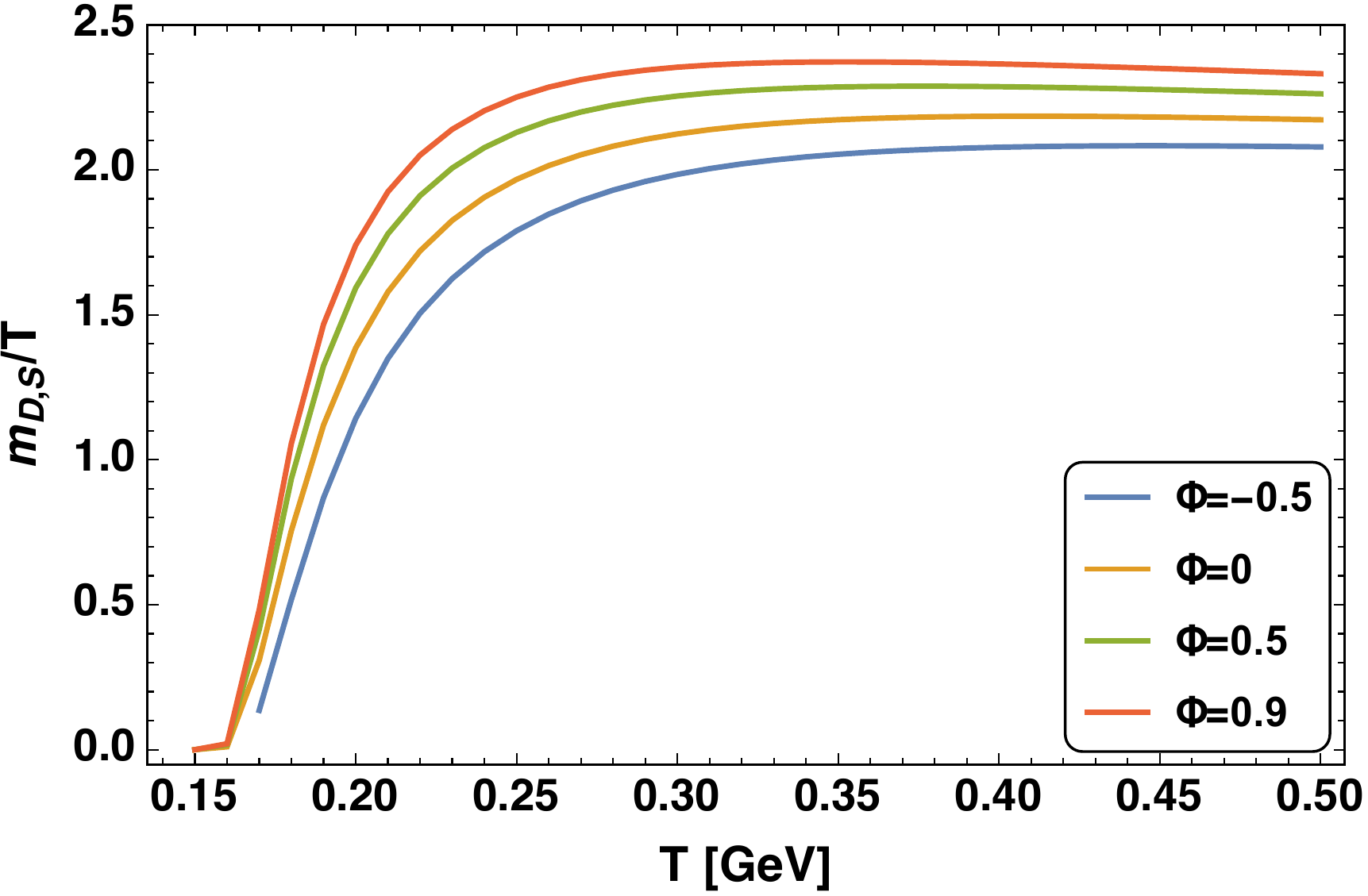}
		}
		\caption{Retarded Debye mass $ m_{D,R} $ (left) and symmetric Debye mass $ m_{D,S} $ (right) normalized with $T$ as a function of temperature for different values of bulk viscous correction $ \Phi $.
		}
		\label{mDRS}
	\end{figure}
	\begin{figure}[tb]
		\begin{center}	
			\includegraphics[width=8.8cm]{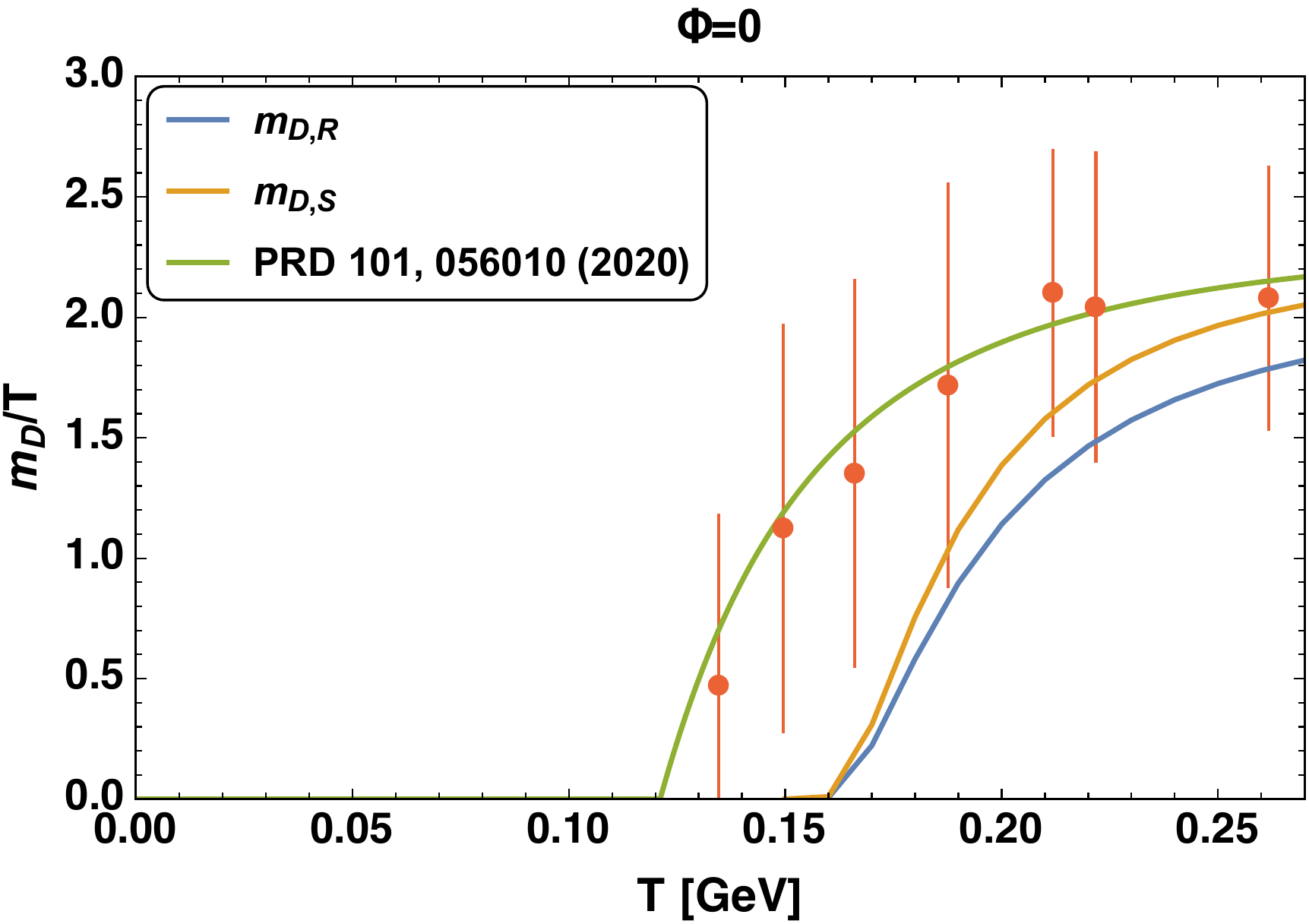}
			\caption{Debye masses ($ m_{D,R} $, $ m_{D,S} $) normalized by $T$ and interpolated Debye mass  together with continuum corrected Debye mass points from ref.~\cite{Lafferty:2019jpr} as a function of $T$ for $ \Phi=0 $. 
			}
			\label{mDcompare}
		\end{center}	
	\end{figure}

	Figure~\ref{mDRS} shows the 
	Debye screening masses, $ m_{D,R} $ and $m_{D,S} $,  
	divided by $T$ as a function of the temperature 
	for different values of $ \Phi $. 
	The Debye masses deviate from the linear dependence 
	on $T$ at lower temperatures, where non-perturbative effects are important. 
	The Debye masses are increasing functions of the bulk viscous parameter $\Phi$. 
	In figure~\ref{mDcompare}, we compare the Debye masses computed here with the Debye mass computed via the 
	fitting of the potential with lattice QCD data~\cite{Lafferty:2019jpr}. %
	The Debye masses $m_{D,R} $ and $m_{D,S} $ 
	show qualitatively similar behaviour as
	those in ref.~\cite{Lafferty:2019jpr}. 
	We find that the symmetric Debye mass is slightly larger than retarded Debye mass in the presence of quasiparticle mass. 
	The effect of chemical potential is quantitatively small
	when $\mu < T$, and hereafter we take $\mu=0$.

	The dielectric permittivity, $ \varepsilon(p) $, 
	can be computed as~\cite{Thakur:2020ifi,Thakur:2016cki}
	\begin{equation} 
		\varepsilon^{-1}(p)=\lim_{p^0  \to 0} {p^2} \bar D_{11} (P)~, 
		\label{ephs}
	\end{equation}
	where $\bar D_{11}(P)$ is the longitudinal component
	of the $11$-part of the resummed gluon propagator.
	The quantity 
	$\bar D_{11}(P)$ 
	can be written as the sum of retarded, advanced and symmetric propagators, 
	\begin{equation}
		\bar D_{11} =\frac{1}{2}\left( \bar D_R  + \bar D_A +\bar D_S  \right) .
		\label{eq:d11}
	\end{equation}
	Using eqs.~(\ref{DRphi}) and (\ref{DSphi}) into eq.~(\ref{ephs}), 
	the dielectric permittivity is written as 
	\begin{equation}
		\varepsilon^{-1}(p) = 
		\frac{p^2}{p^2 + \widetilde{m}^2_{D,R}}
		- i  
		\frac{\pi T p  \, 
			\widetilde{m}^2_{D,S}  
		}{ (p^2 + \widetilde{m}^2_{D,R})^2} . 
		\label{eq:epsilonphi}
	\end{equation}
	In the limit of vanishing mass and bulk viscous correction, 
	both the Debye masses
	$\widetilde{m}_{D,R}$ and $\widetilde{m}_{D,S}$ 
	approaches to the equilibrium Debye mass $m_D$ and  the equilibrium expression is reproduced. 
	The dielectric permittivity 
	(\ref{eq:epsilonphi}) will be used 
	in the computation of the in-medium heavy quark potential. 
	%%%%%%%%%%%%%%%%%%%%%%%%%%%%%%%%%%%%%

	\section{Heavy quarkonia in a bulk viscous medium}
	\label{sec:quarkonia}
	In this section, we discuss the effects of bulk viscous correction on the physical properties of heavy quarkonia. 
	First, we compute the in-medium heavy quark potential with the modified permittivity computed in the previous section. 
	Based on the potential, we compute the spectral functions, 
	from which the physical properties of quarkonia can be extracted. 
	
	\subsection{In-medium heavy quark potential}\label{sec:potbulk}
	
	In this subsection, we discuss how 
	the heavy quark potential is modified 
	in a bulk viscous medium. 
	Following previous approaches~\cite{Thakur:2013nia,Thakur:2020ifi} based on the linear response theory,  
	we compute the in-medium heavy-quark potential
	by modifying the Cornell potential 
	using the dielectric permittivity  $\varepsilon(p)$ 
	as $V({\bf p}) = V_{\rm Cornell}(p) / \varepsilon(p)$, 
	where 
	$V_{\rm Cornell}(p)$ is the Cornell potential in the momentum space. 
	The Fourier transform of $V({\bf p})$ is given by  
	\begin{eqnarray}
		\label{defn}
		V(r,T)&=&\int \frac{d^3\mathbf p}{{(2\pi)}^{3/2}}
		(e^{i\mathbf{p} \cdot \mathbf{r}}-1)~\frac{V_{\rm Cornell} (p)}{\varepsilon(p)} ~,
		\label{potdef}
	\end{eqnarray}
	The Cornell potential is parametrized as 
	\begin{equation}
		V_{\rm Cornell} (r)=-\frac{\alpha}{r}+\sigma r, 
		\label{cor-r}
	\end{equation}
	where $ \alpha = C_{F}~ g^2/4\pi $ 
	with $C_{F}=(N_c^2-1)/2N_c$ 
	is the strong coupling constant and
	$ \sigma $ is the string tension. 
	Here  we take, $ \alpha=0.513 $ and 
	$ \sigma =(0.412 ~{\rm GeV})^2$~\cite{Lafferty:2019jpr}. 
	The Fourier transform of eq.~\eqref{cor-r} is given by 
	\begin{equation}
		V_{\rm Cornell} (p)=-\sqrt{(2/\pi)} \frac{\alpha}{p^2}-\frac{4\sigma}{\sqrt{2 \pi} p^4}. 
		\label{eq:vcornell}
	\end{equation}
	After substituting 
	eq.~(\ref{eq:epsilonphi}) into eq.~(\ref{potdef}), we obtain 
	an in-medium complex heavy quark potential. 
	Its real part reads 
	\begin{equation}
		\begin{split} 
			{\rm Re}\, V(r,T,\Phi)
			&= 
			\int \frac{d^3\mathbf p}{{(2\pi)}^{3/2}}
			(e^{i\mathbf{p} \cdot \mathbf{r}}-1)
			V_{\rm Cornell} (p)
			{\rm Re}\, \varepsilon^{-1}(p)  \\
			&= 
			-\alpha \, 
			\widetilde{m}_{D,R}
			\left(\frac{e^{-\widetilde{m}_{D,R}\, r}}{\widetilde{m}_{D,R}\, r}+1\right) 
			+ 
			\frac{2\sigma}{
				\widetilde{m}_{D,R} 
			}\left(\frac{e^{-\widetilde{m}_{D,R}\,{r}}-1}{\widetilde{m}_{D,R}\,r}+1\right) ,
			\label{eq:pot-realphi}
		\end{split}
	\end{equation}
	In addition to eq.~\eqref{eq:pot-realphi}, 
	we add a constant $c$, whose value we take $ c = -0.161 $~GeV~\cite{Lafferty:2019jpr}. 
	The first term
	in eq.~\eqref{eq:pot-realphi} 
	is the perturbative Coulombic part with a screening, 
	and the second term represents non-perturbative contributions. 
	In the small distance limit ($r \to 0$), 
	eq.~(\ref{eq:pot-realphi}) approaches to the Cornell potential. 
	The bulk viscous correction enters 
	through the modification of the Debye mass,  
	$m_{D, R} \to \widetilde{m}_{D, R}$.

	\begin{figure}[tb]
		\subfigure{
			\hspace{-0mm}
			\includegraphics[width=7.8cm]{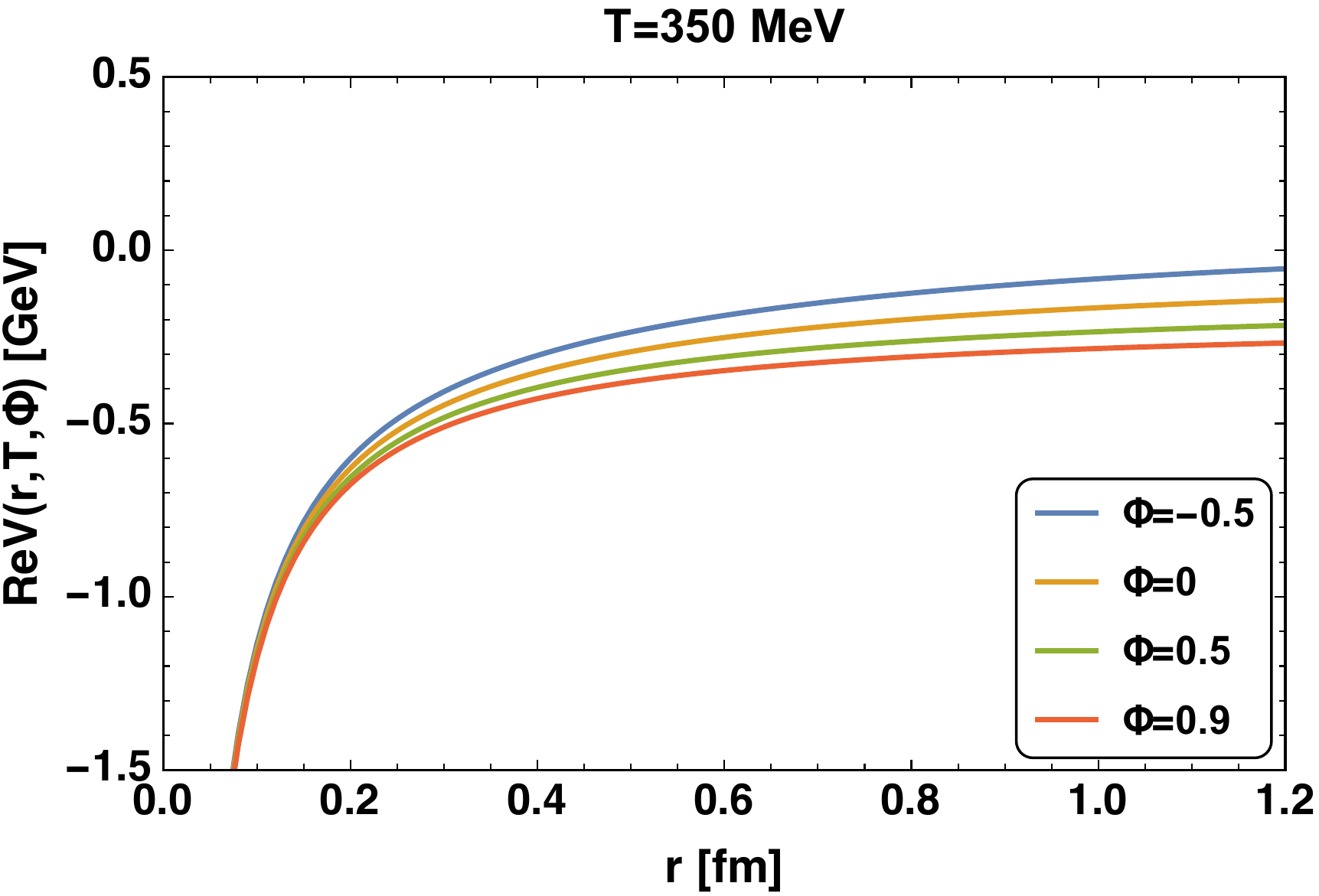}}
		\subfigure{
			\includegraphics[width=7.6cm]{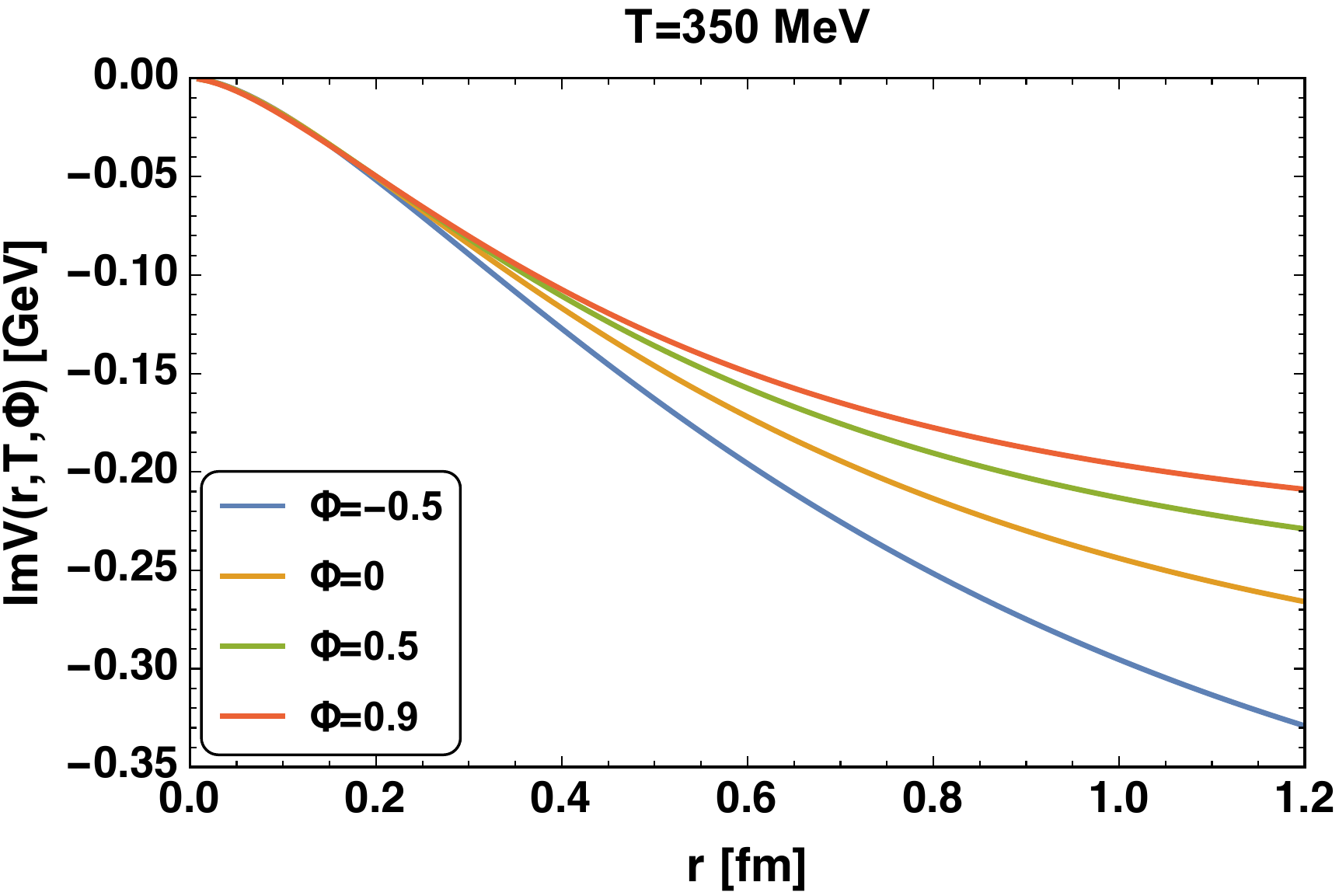}}
		\caption{
			Real (left) and imaginary (right) part of the potential as a function of $ r  $  for different values of $ \Phi $ at $ T=350 $ MeV.
		}
		\label{fig:potT350}
	\end{figure}
	The imaginary part of the potential is written as 
	\begin{equation}
		\begin{split} 
			{\rm Im}\, V(r,T,\Phi)&=\int \frac{d^3\mathbf p}{{(2\pi)}^{3/2}}
			(e^{i\mathbf{p} \cdot \mathbf{r}}-1)
			V_{\rm Cornell}(p) {\rm Im}\, \varepsilon^{-1}  (p)  
			\\
			&= 
			-\alpha \lambda T 
			\, \phi_2  (\widetilde{m}_{D,R}\, r)
			- 
			\frac{ 
				2\sigma T \lambda
			} {
				%m_D^2
				\widetilde{m}^2_{D, R} 
			}
			\, 
			\chi( \widetilde{m}_{D, R}\, r ),
			%	&\equiv
			%	{\rm Im}\, V_{\rm HTL}(r)
			% 		+ 
			% 		{\rm Im}\, V_{\rm string}(r),
			\label{eq:pot-imaginary1}
		\end{split}
	\end{equation}
	%where the first term is from the perturbative HTL contribution, 
	%and the second term is the string-like contribution. 
	% 
	where the dimensionless parameter $\lambda$ is defined as 
	\begin{equation}
		\lambda = 
		\lambda ( \Phi, \tilde \mu)
		\equiv 
		\frac{ \widetilde m^2_{D, S} } { \widetilde m^2_{D, R} } . 
		\label{eq:def-lambda2}
	\end{equation}
	The functions $\phi_n(x)$ and $\chi(x)$ are defined as
	\begin{equation}
		\phi_n (x)
		\equiv 
		2 \int_{0}^{\infty} dz
		\frac{z}{(z^{2}+1)^{n}}\left[1-\frac{\sin(x z)}{x z}\right] ,
	\end{equation}
	\begin{equation}
		\chi (x) 
		\equiv
		2 \int_{0}^{\infty}  \frac{dz}{z (z^{2}+1)^{2}}
		\left[1-\frac{\sin(x z)}{x z}\right].
		\label{eq:chi}
	\end{equation}
	The function $\chi  (x)$ is monotonically increasing with $\chi(0) =0$, 
	and is logarithmically divergent at large $x$. 
	We regularize this by modifying the following part 
	in eq.~\eqref{eq:chi} as 
	\begin{equation}
		\frac{1}{z(z^{2}+1)^{2}}\rightarrow \frac{1}{\sqrt{z^2+\Delta^2}~(z^{2}+1)^{2}},
	\end{equation}
	with $ \Delta\simeq 3.0369 $~\cite{Lafferty:2019jpr}. 
	
	In figure~\ref{fig:potT350}, 
	we plot the real and imaginary part of the potential. 
	We find that the real part of the potential becomes more flattened for a larger $\Phi$, due to an increased screening. %
	As for the imaginary part, 
	its magnitude is suppressed at larger values of $r$ 
	for $\Phi>0$. 

	Let us note the difference in the behaviour of the potential 
	of the current and previous \cite{Thakur:2020ifi} work. 
	%
	%The behaviour of real and imaginary part of the potential in the presence of quasiparticle mass are not the same as that of the real and imaginary part of the potential computed in our previous work \cite{Thakur:2020ifi} for the massless case. 
	%
	The main difference is that we have introduced quasiparticle masses in the current work, while the thermal particles were massless in the previous one. 
	This results in the different temperature dependence
	of Debye masses. 
	While the Debye masses are always linear in temperature 
	in the previous work, 
	they show non-linear dependence at lower temperatures 
	in the current work (see Fig.~\eqref{mDRS}) and
	is suppressed in this region. 
	Thus, 
	the screening effect in the potential 
	is smaller in the present work, 
	which leads to larger values of binding energies as we 
	see later.

	%In the present case, Debye masses ($ m_{D,R},m_{D,S} $ ) show non-linear dependence as a function of temperature which results in the decrease in screening at lower temperatures. 
	%On the other hand, in our previous work, Debye screening masses show linear dependence with temperature, hence the screening is more compared to the present case.
	%Those changes of the shape of the potential in the presence of quasiparticle mass are 
	%going to affect the properties 
	%of quarkonia, as we see in the following subsection. 

	\subsection{In-medium spectral functions of heavy quarkonia}\label{sec:spectral function} 
	
	We here discuss the spectral functions of heavy quarkonia in a bulk viscous medium. 
	From the spectral functions, we can 
	extract physical properties of heavy quarkonia 
	such as binding energies and decay widths. 
	For the computation of the spectral functions, 
	we employ the Fourier space method developed in ref.~\cite{Burnier:2007qm}. 
	In this method, 
	the quantity of interest is the 
	unequal-time point-split meson-meson correlator, 
	$G^{>} (t;{\bf r},{\bf r'})$. 
	The vector-channel spectral function can be obtained 
	from the Fourier transform of the correlator as  
	\begin{equation}
		\rho^V({\omega})=\lim_{{\bf r},{\bf r'}\rightarrow 0}\frac{1}{2}\tilde{G}(\omega;{\bf r},{\bf r'}),
	\end{equation}
	where 
	\begin{equation}
		\tilde{G}(\omega;{\bf r},{\bf r'})=\int_{-\infty}^{\infty}dt e^{i\omega t}G^{>} (t;{\bf r},{\bf r'}). 
	\end{equation}
	The correlator is shown to satisfy 
	the following Schr\"odinger equation, 
	\begin{equation}
		\left[{\hat{H} }\mp i | {\rm Im} V(r,T,\Phi)|\right]G^{>} (t;{\bf r},{\bf r'})=i\partial_t G^{>} (t;{\bf r},{\bf r'}), 
		\label{schreq}
	\end{equation}
	where
	\begin{equation}
		\hat{H}=2m_Q-\frac{\nabla^{2}_{r}}{m_Q}+\frac{l(l+1)}{m_Qr^2}
		+{\rm Re }\, V(r,T,\Phi).
	\end{equation}
	In solving this, we use the potential derived in the previous subsection. 
	Further details of the computation of the spectral functions from eq.~(\ref{schreq}) can be found in ref.~\cite{Burnier:2007qm}.

	\begin{figure}[tb]
		\subfigure{
			\hspace{-8mm}
			\includegraphics[width=7.8cm]{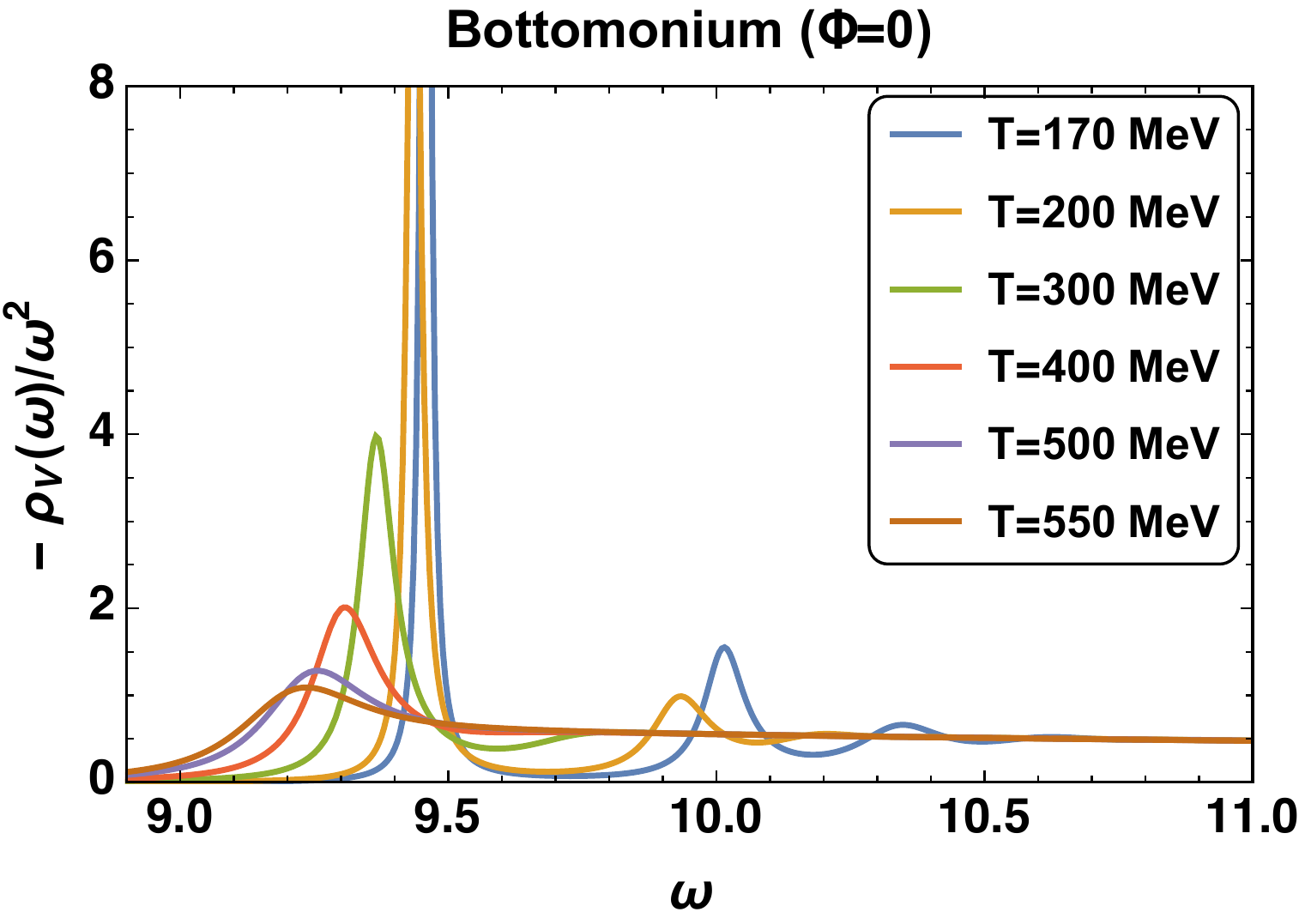}
		}
		\subfigure{
			\includegraphics[width=7.8cm]{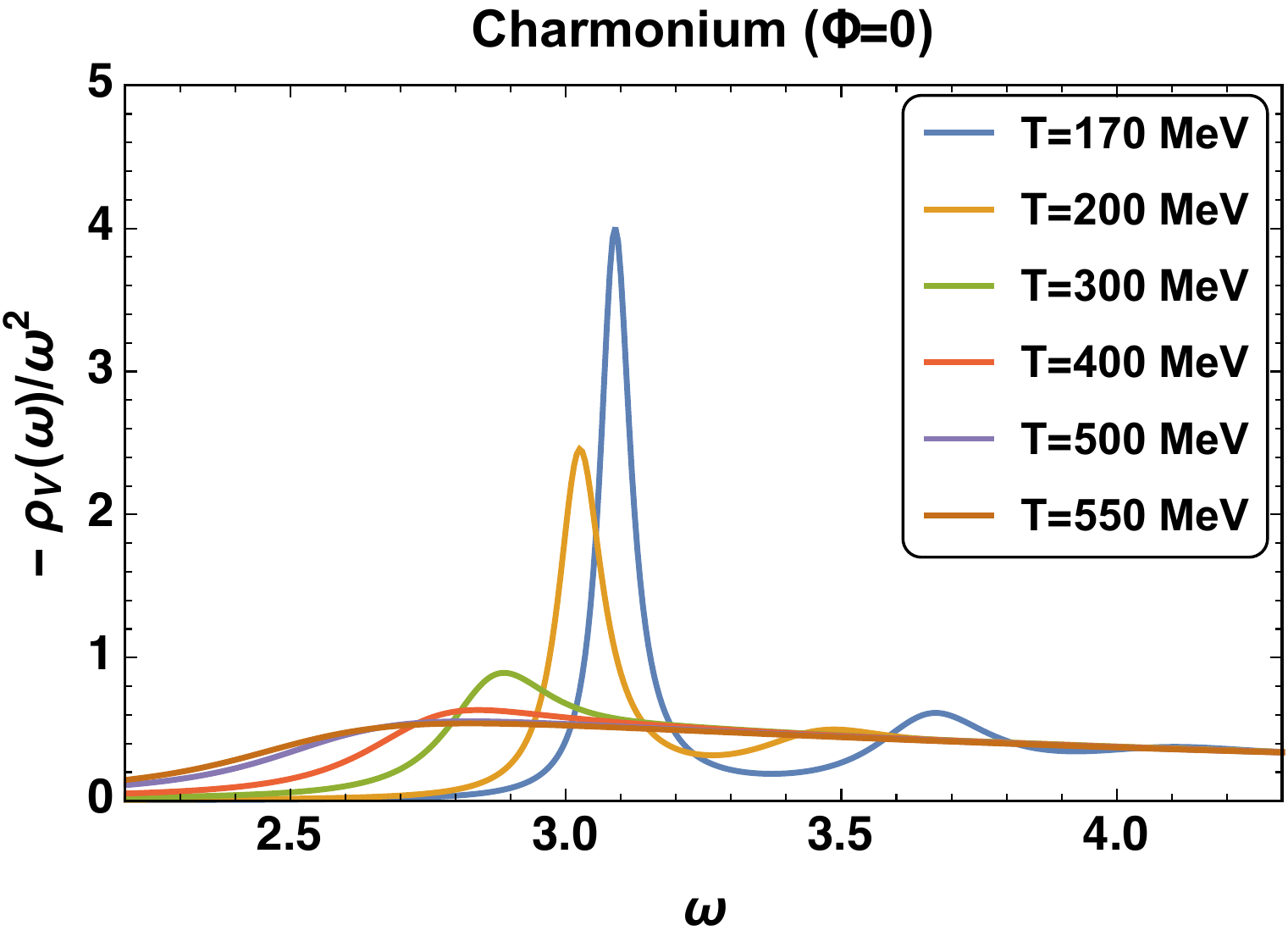}
		}
		\caption{
			S-wave spectral functions for vector channel bottomonium (left) and charmonium (right) for different  values of temperature at $ \Phi=0 $.
		}
		\label{SFbotcharm}
	\end{figure}
	%
	%%%%%%%%%%%%%%%%%%%%%%%%%%%%%%%%%%%%
	%%%%%%%%%%%%%%%%Fig170%%%%%%%%%%%%%%%%%
	\begin{figure}[tb]
		\subfigure{
			\hspace{-8mm}
			\includegraphics[width=7.8cm]{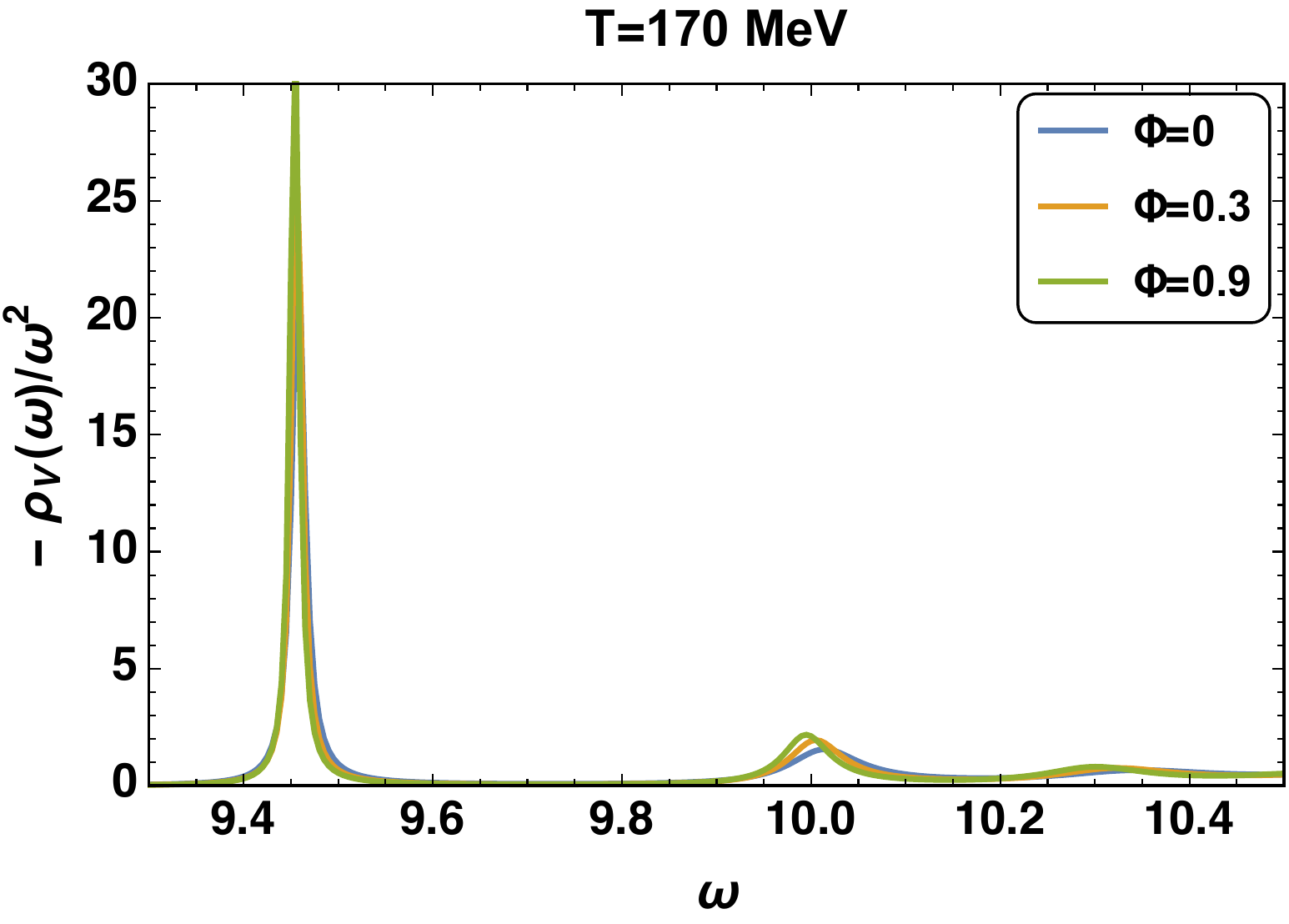}
		}
		\subfigure{
			\includegraphics[width=7.8cm]{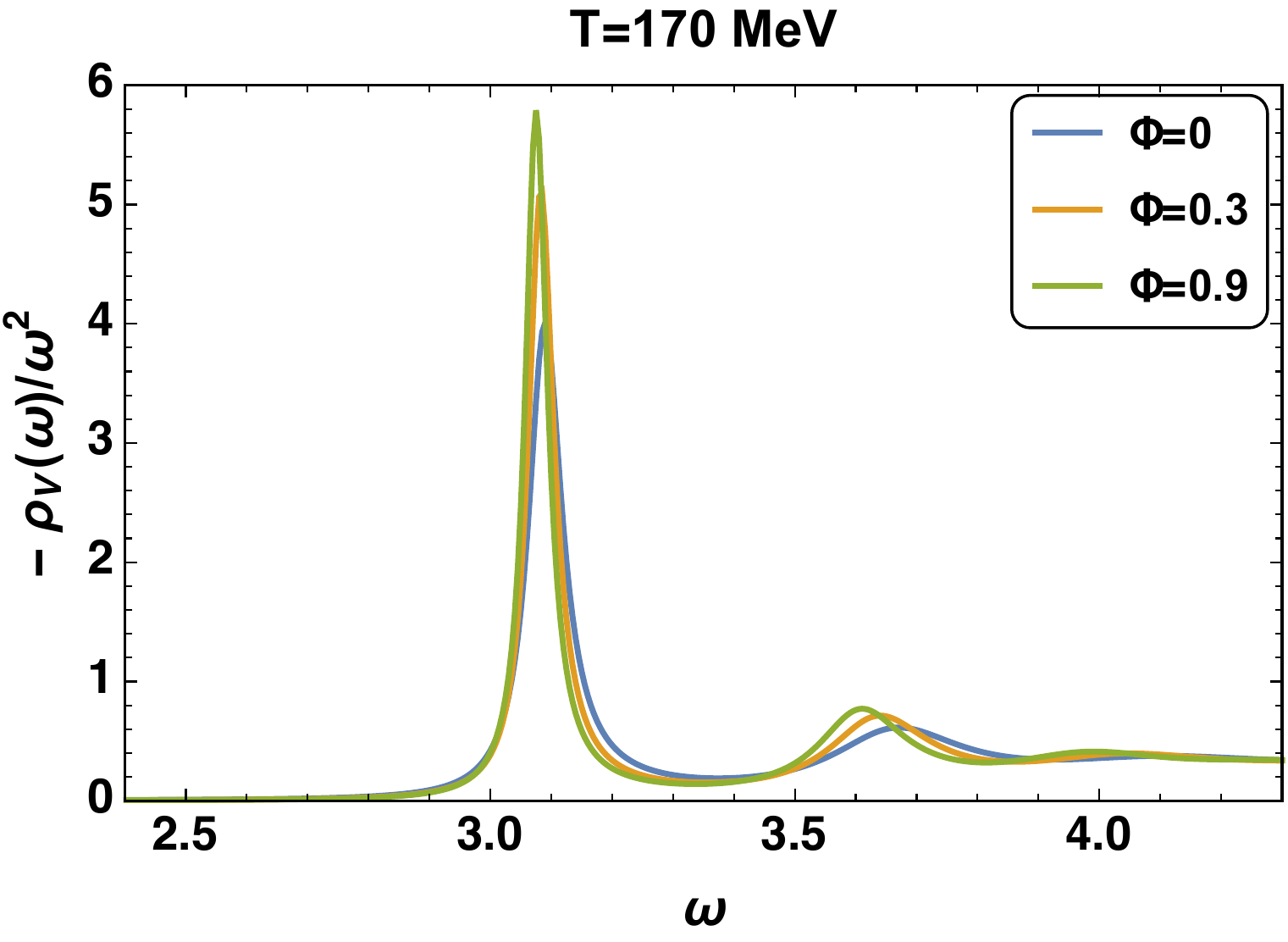}
		}
		\caption{ S-wave spectral functions for vector channel bottomonium (left) and charmonium (right) for different values of $ \Phi $ at $T=170$ MeV.
		}
		\label{SFT170phi}
	\end{figure}

	In figure~\ref{SFbotcharm}, we plot the spectral functions 
	of bottomonium and charmonium for different values of $T$ at $\Phi=0$. 
	The spectral peaks 
	shift towards lower values of $\omega$ 
	at higher temperatures, 
	which is consistent with the results reported in recent works~\cite{Lafferty:2019jpr,Burnier:2015tda}.
	The broadening of peaks occurs 
	because of the imaginary part of the potential, 
	whose magnitude increases at higher temperatures.

\begin{figure}[tb]
		\subfigure{
			\hspace{-8mm}
			\includegraphics[width=7.8cm]{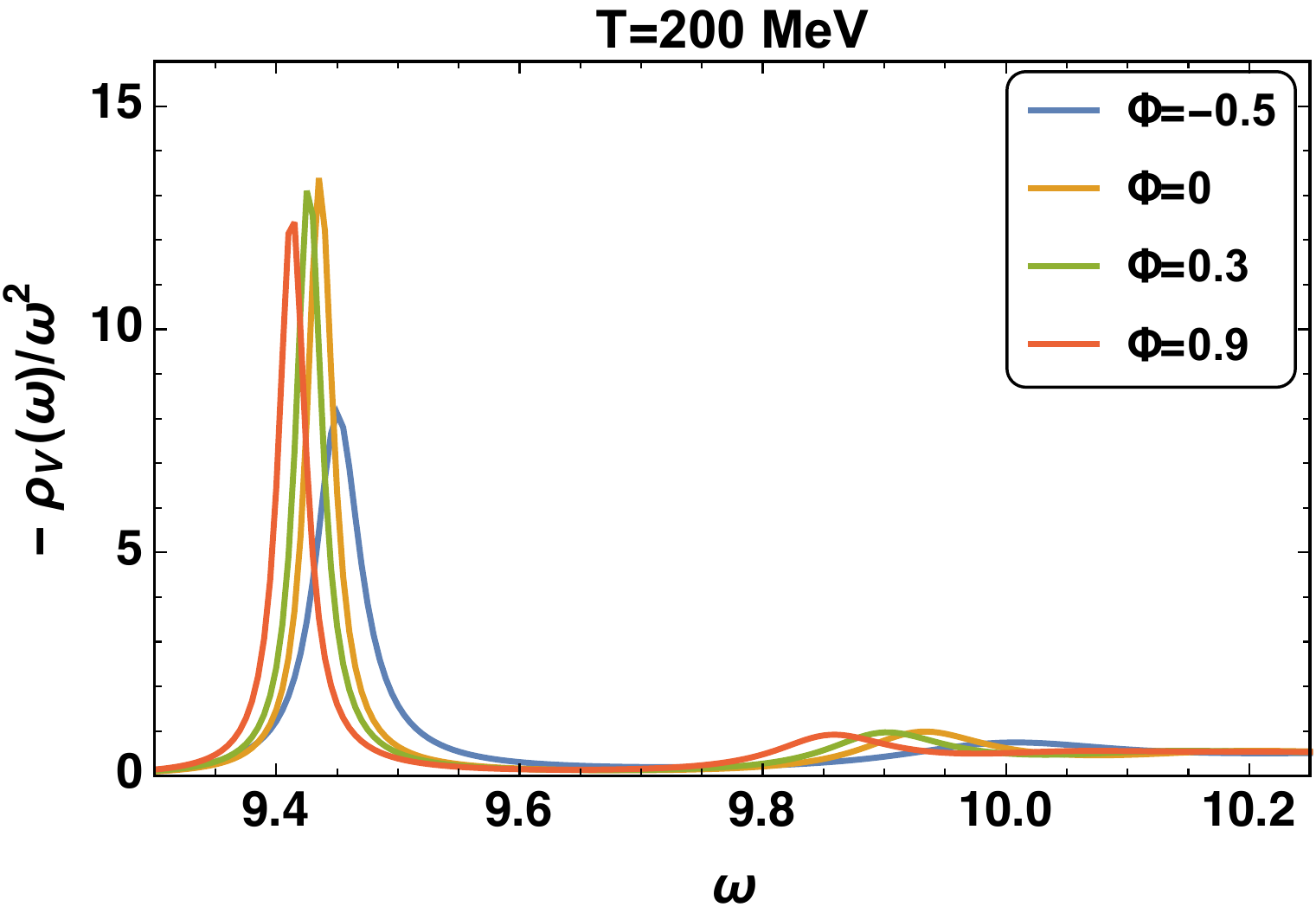}
		}
		\subfigure{
			\includegraphics[width=7.8cm]{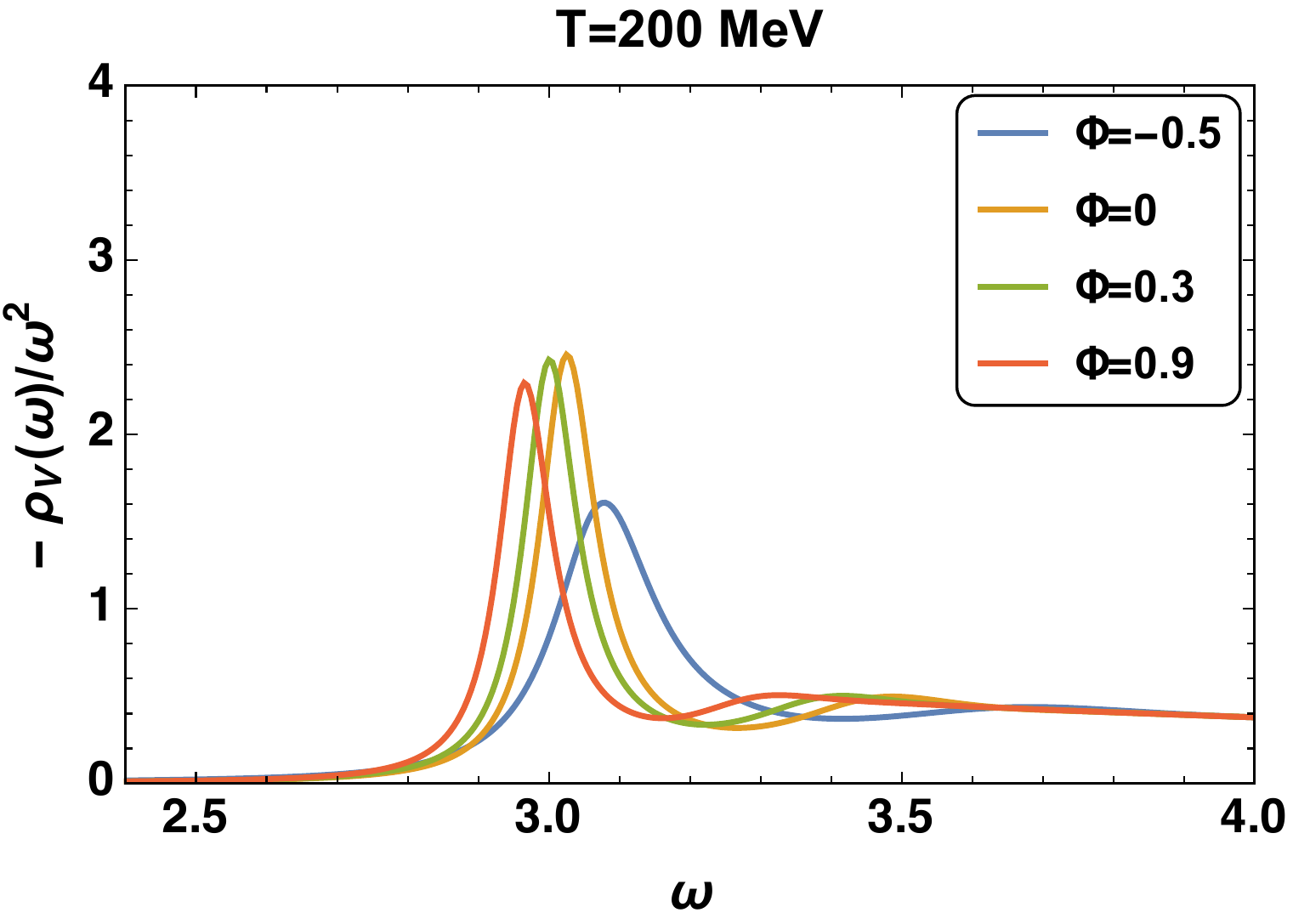}
		}
		\caption{
			S-wave	spectral functions for vector channel  bottomonium (left) and charmonium (right) for different values of $ \Phi $ at $T=200 $ MeV.
		}
		\label{SFT200phi}
	\end{figure}

\begin{figure}[tb]
		\subfigure{
			\hspace{-8mm}
			\includegraphics[width=7.8cm]{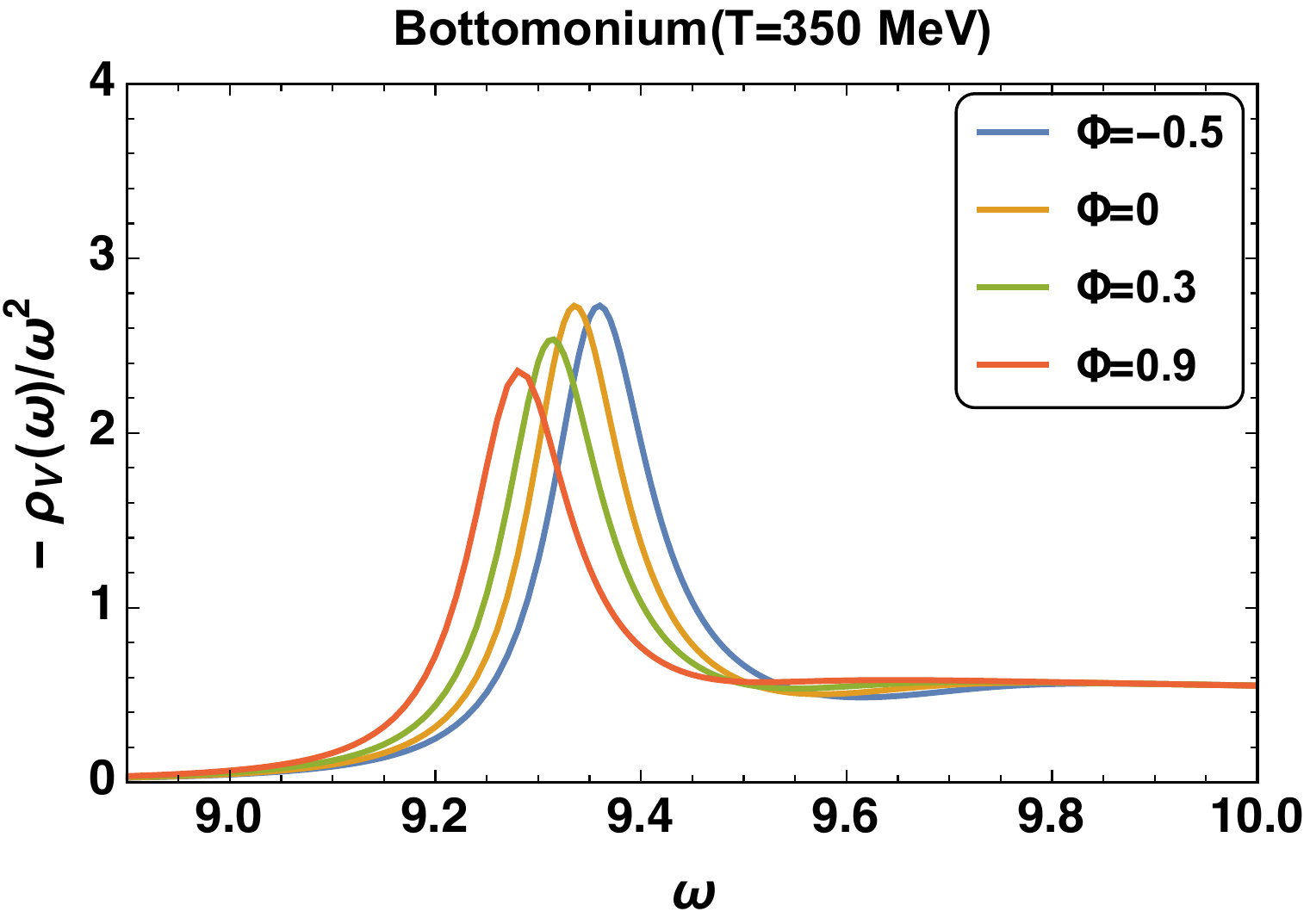}
		}
		\subfigure{
			\includegraphics[width=7.8cm]{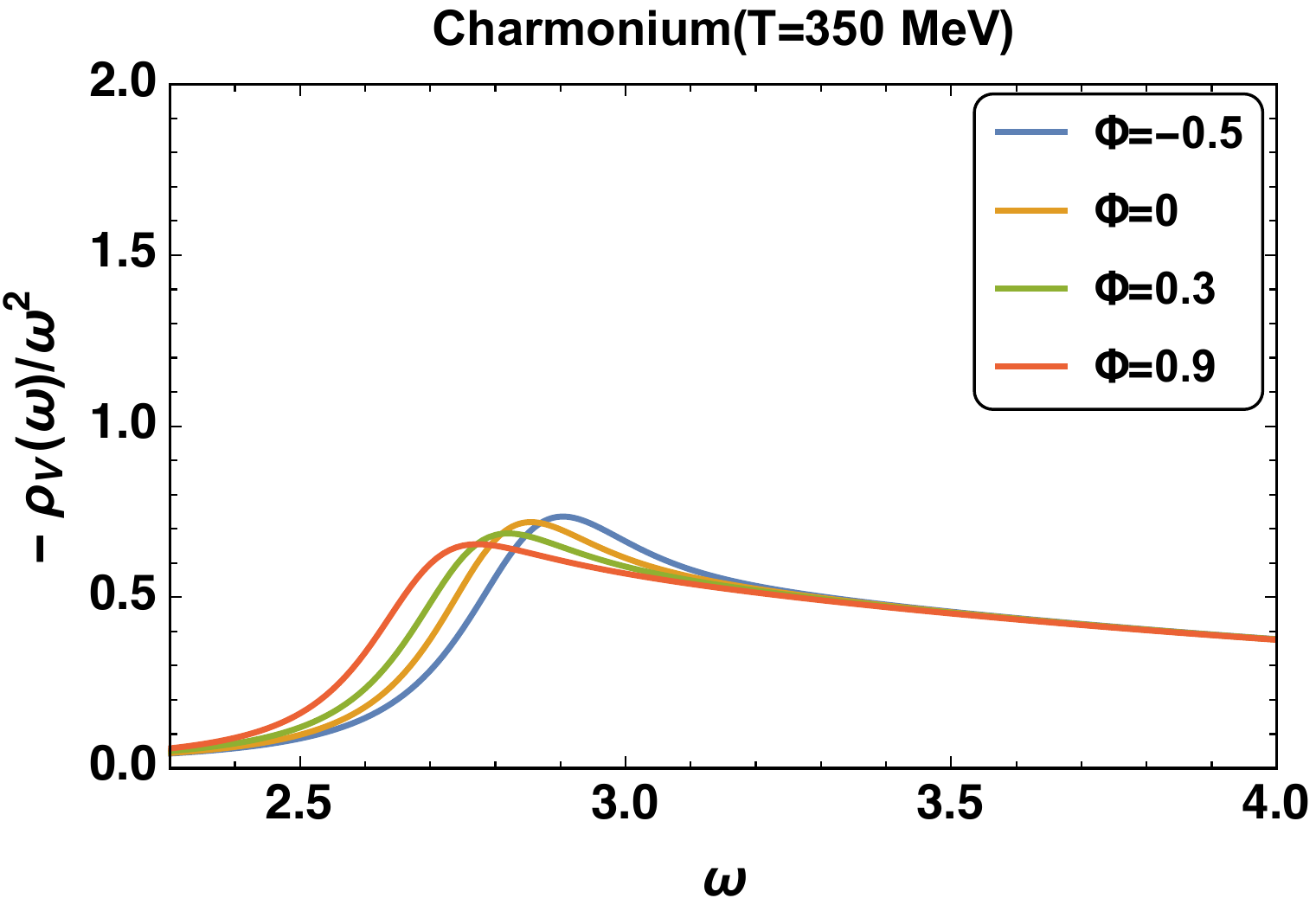}
		}
		\caption{
			S-wave spectral functions for vector channel bottomonium (left) and charmonium (right) for different values of $ \Phi $ at $ T=350 $ MeV.
		}
		\label{SFT350phi}
	\end{figure}

\begin{figure}[tb]
		\subfigure{
			\hspace{-8mm}
			\includegraphics[width=7.8cm]{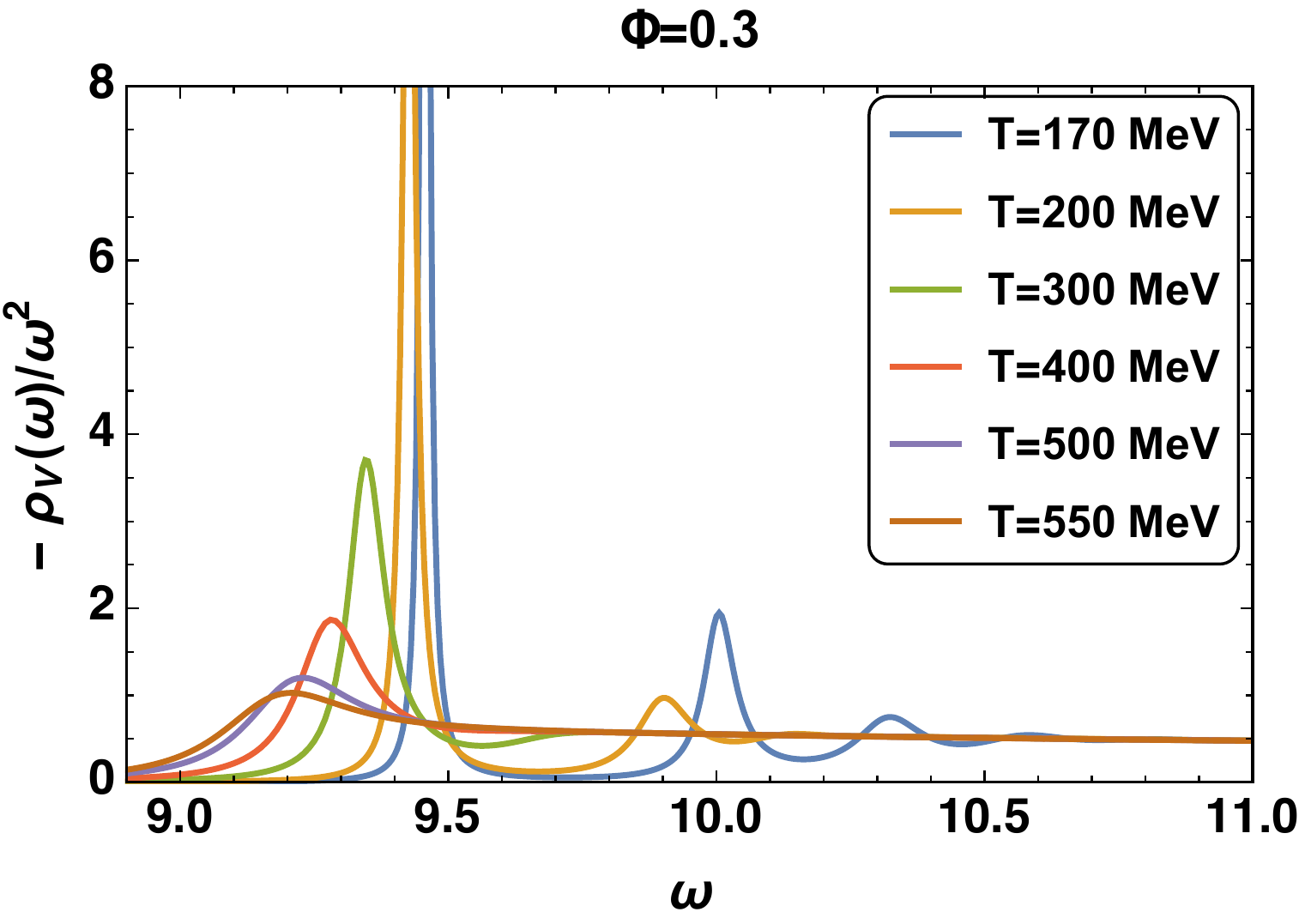}
		}
		\subfigure{
			\includegraphics[width=7.8cm]{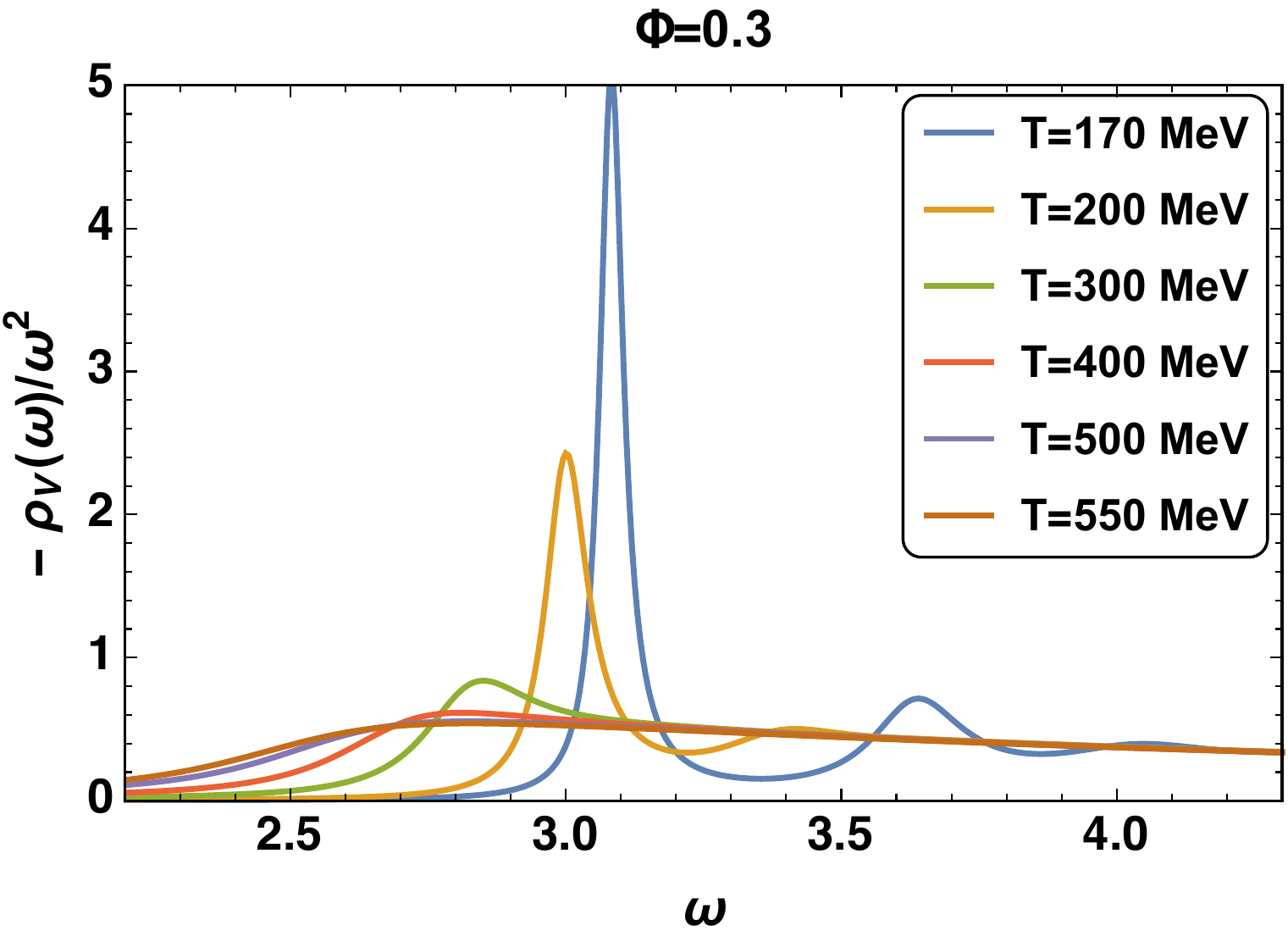}
		}
		\caption{
			S-wave	spectral functions for vector channel bottomonium (left) and  charmonium (right) for different values of $ T $ at $ \Phi =0.3$.
		}
		\label{SFphi03}
	\end{figure}

	Figures~\ref{SFT170phi}, \ref{SFT200phi} and  \ref{SFT350phi} show the spectral functions of bottomonium and charmonium
	at $ T=170, 200, 350$ MeV for different values of $\Phi$.
	Figure~\ref{SFphi03} shows the spectral functions  
	for different values of temperature for $ \Phi=0.3 $. 
	Spectrum peaks get shifted towards lower values of $ \omega $ with the increase in $\Phi$.  
	Above a certain temperature, the spectral functions of charmonium  and bottomonium states show no resonance like structure. 
	%
	%
	%
	%%%%%%%%%%%%%%%%%%%%%%%%%%%%%%%%%%%%
	%%%%%%%%%%%%%%%%Fig%%%%%%%%%%%%%%%%%
	%%%%%%%%%%%%%%%%Fig350%%%%%%%%%%%%%
	%%%%%%%%%%%%%%%%%%%%%%%%%%%%%%%%%%%%%%%

	To quantify the physical properties of each state, 
	we fit each peak of the in-medium spectral functions 
	with the following skewed Breit-Wigner form~\cite{Burnier:2015tda,Burnier:2016kqm,Lafferty:2019jpr}, 
	\begin{equation}
		\rho(\omega\approx E)=C\frac{(\Gamma/2)^2}{(\Gamma/2)^2+(\omega-E)^2}+2\delta\frac{(\omega-E)\Gamma/2}{(\Gamma/2)^2+(\omega-E)^2}+A_{1}+A_{2}(\omega-E)+O(\delta^2),
		\label{BW}
	\end{equation}
	where $E$ is the energy of the resonance, 
	$\Gamma$ is its width, 
	$\delta$ is the phase shift, 
	$A_1$ and $A_2$ are parameters 
	to account for structures 
	unrelated to the peak of interest. 
	For each state, 
	we obtain 
	the resonance mass and 
	the decay width, 
	and we repeat this process 
	for several different values of $T$ and $\Phi$.

	\begin{figure}[tb]
		\subfigure{
			\hspace{-8mm}
			\includegraphics[width=7.8cm]{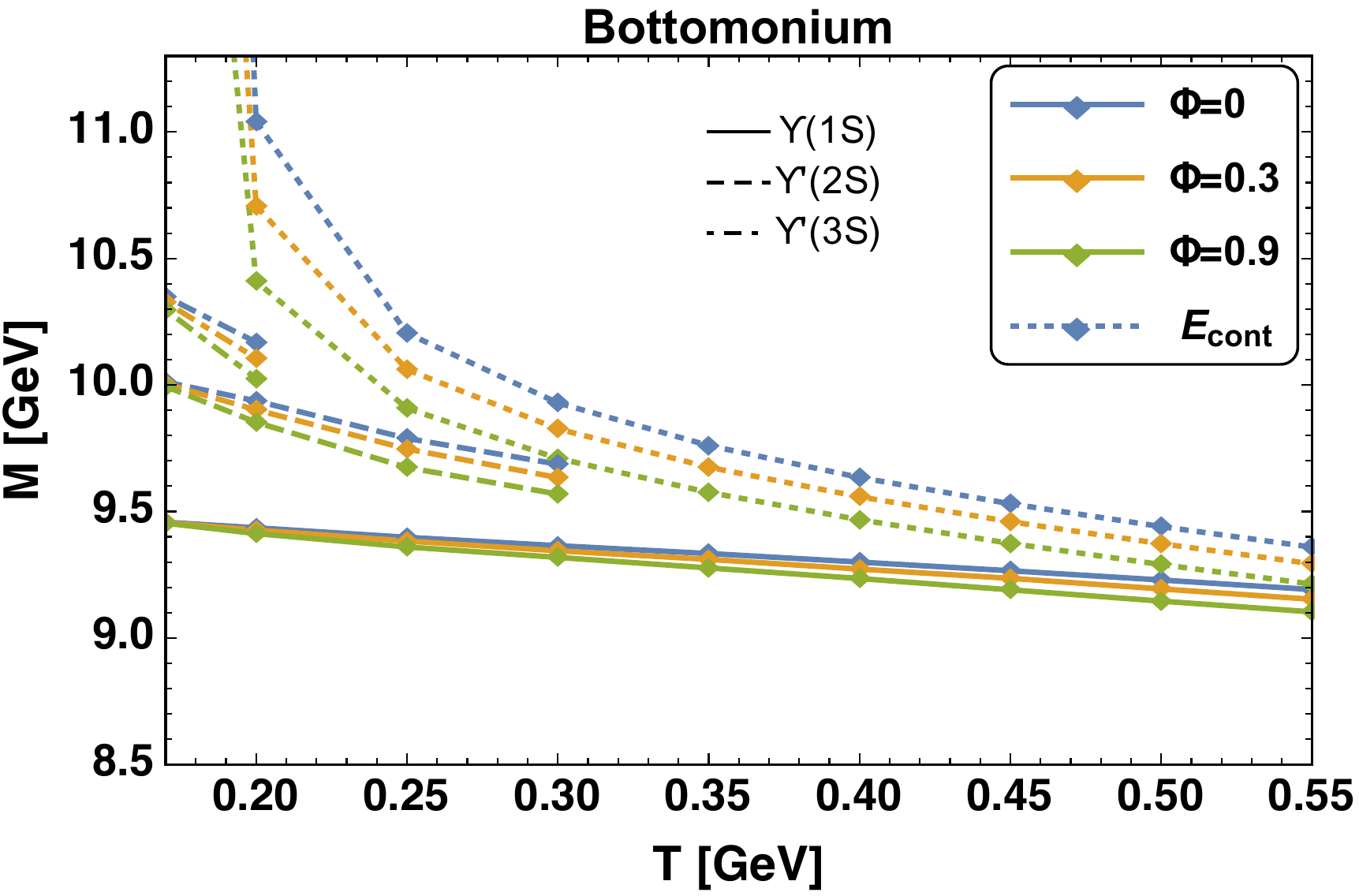}
		}
		\subfigure{
			\includegraphics[width=7.8cm]{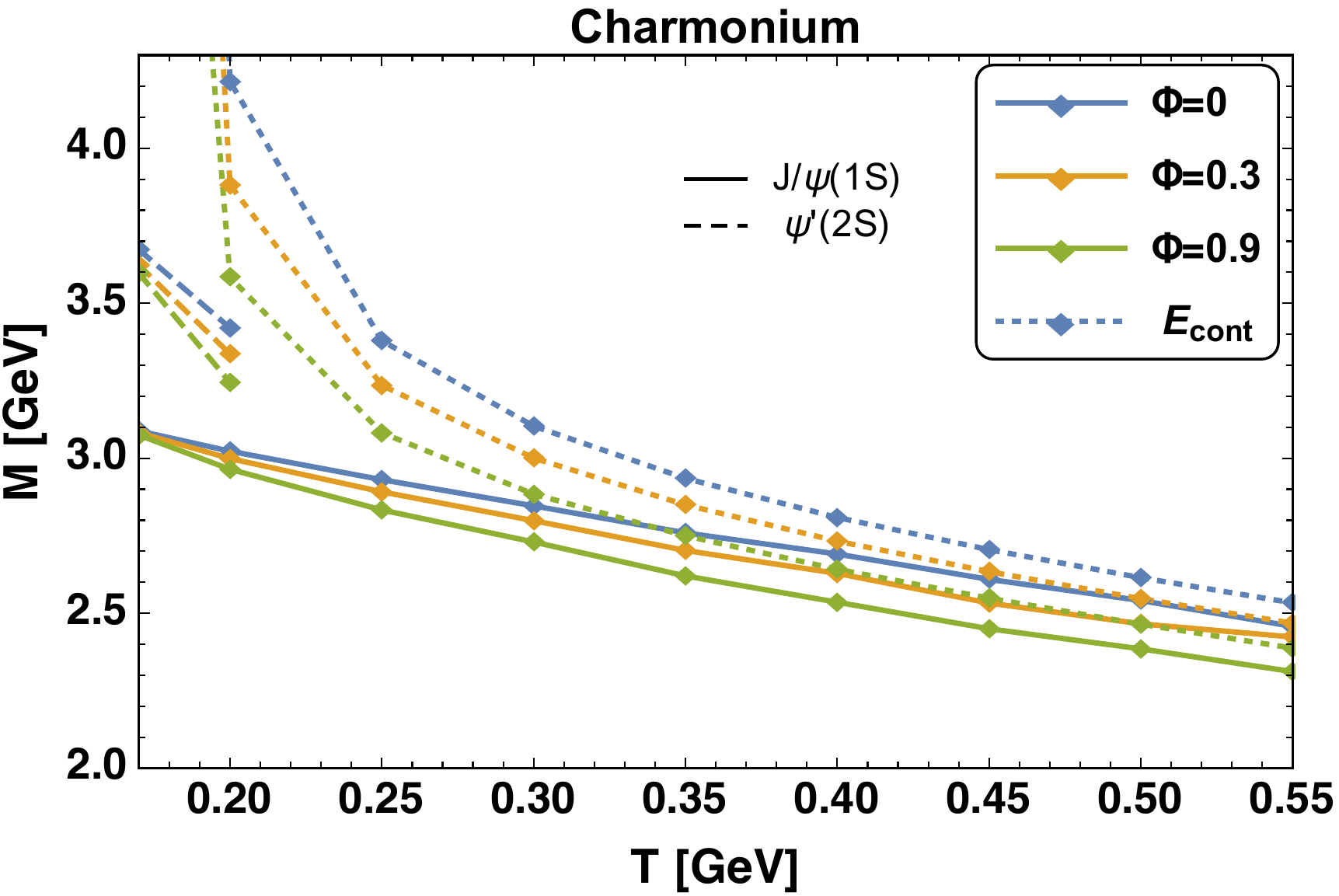}
		}
		\caption{
			The resonance masses of 
			bottomonium (left) and charmonium (right) bound states for different values of $ \Phi $. 
			The continuum threshold energy is defined by 
			$E_{\rm cont} = 2m_{Q}+{\rm ReV}(r\rightarrow \infty ) $.
		}
		\label{MEcont}
	\end{figure}
	Figure~\ref{MEcont} shows the in-medium masses, $M$, 
	of bottomonium states, i.e.,  $  \Upsilon (1S)$ (solid line), $ \Upsilon^{\prime} (2S)$ (dashed line), $ \Upsilon^{\prime} (3S)$ (dot-dashed line) and  charmonium states, i.e., $  J/\psi(1S)$ (solid line), $ \psi^{\prime}(2S)$ (dashed line) as a function of $T$ for different values of $ \Phi $. 
	We also plot the continuum threshold energy (dotted lines), $ E_{\rm cont} $, defined by 
	$ 2m_{Q}+{\rm ReV}(r\rightarrow \infty )$. 
	The resonance mass of the each state and 
	the continuum threshold energy show a monotonous 
	decrease as a function of $T$, 
	which is consistent with the previous results~\cite{Burnier:2015tda,Lafferty:2019jpr}.   
	The presence of bulk viscous corrections
	further pushes down the 
	in-medium masses and threshold energy.

	\begin{figure}[tb]
		\subfigure{
			\hspace{-8mm}
			\includegraphics[width=7.8cm]{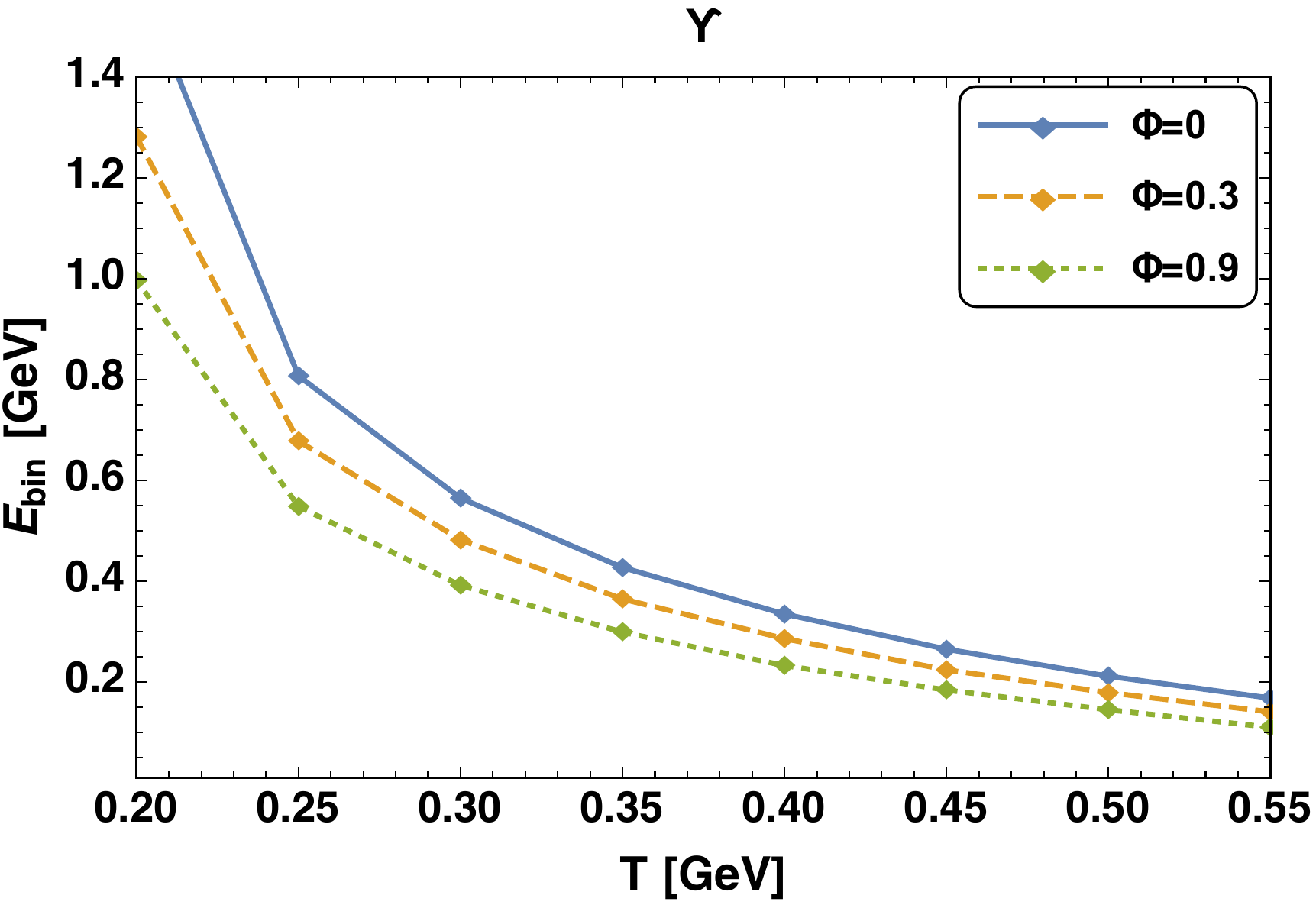}
		}
		\subfigure{
			\includegraphics[width=7.8cm]{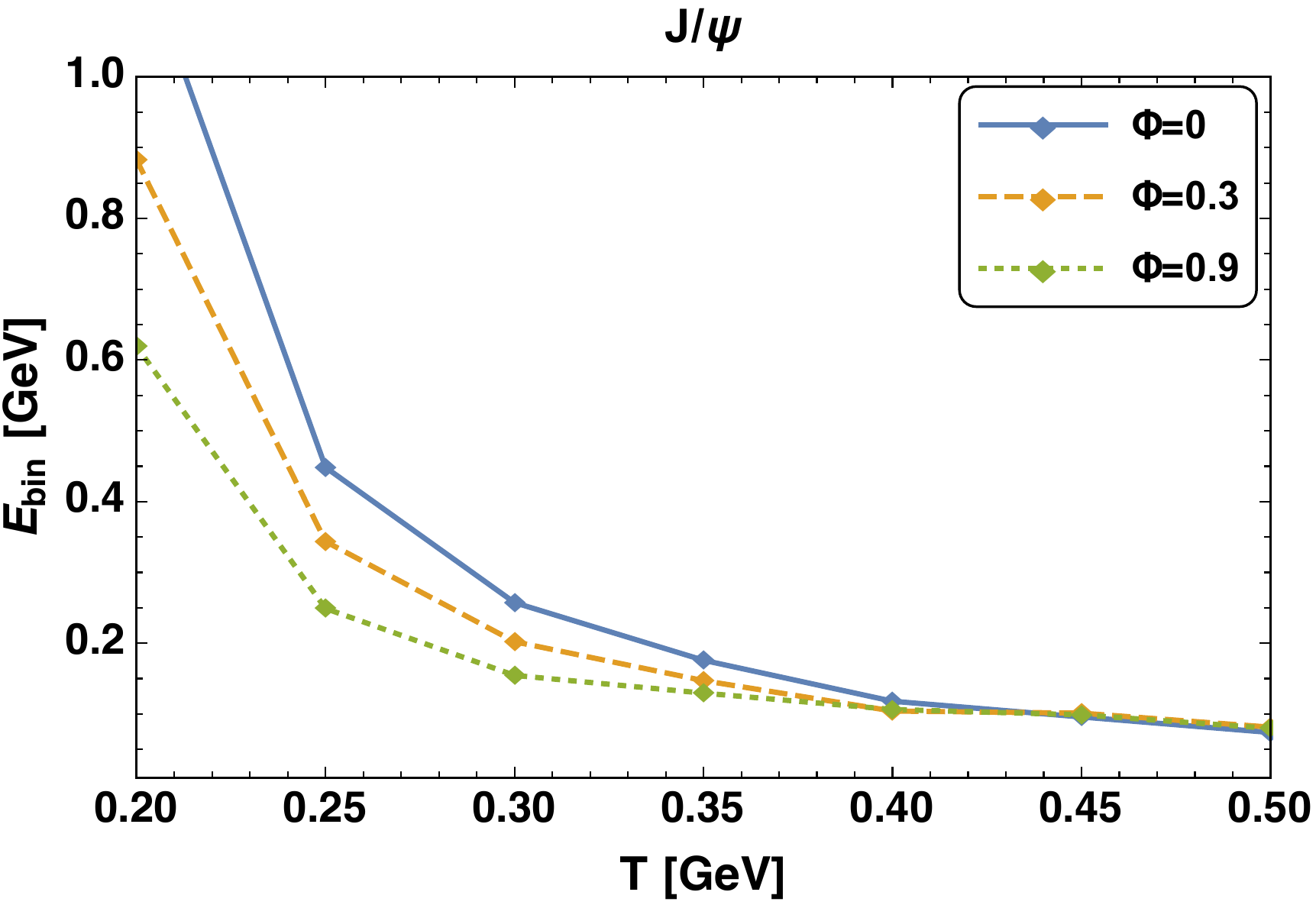}
		}
		\caption{
			The binding energies of the $\Upsilon (1S)$ (left) and $J/\psi (1S)$ (right) for different values of $ \Phi $.
		}
		\label{Ebin}
	\end{figure}

	Figure~\ref{Ebin} shows the binding energies of the 
	$\Upsilon(1S)$ and $J/\psi (1S)$
	as a function of temperature for different values of $\Phi$. 
	The binding energies are decreasing functions of both $T$ and $\Phi$, which is consistent to our previous results~\cite{Thakur:2020ifi}.  
	Quantitatively, the values of the binding energies 
	are higher in the present work, 
	because of the reduced screening effect 
	from the introduction of quasiparticle masses.

\begin{figure}[tb]
		\subfigure{
			\hspace{-8mm}
			\includegraphics[width=7.8cm]{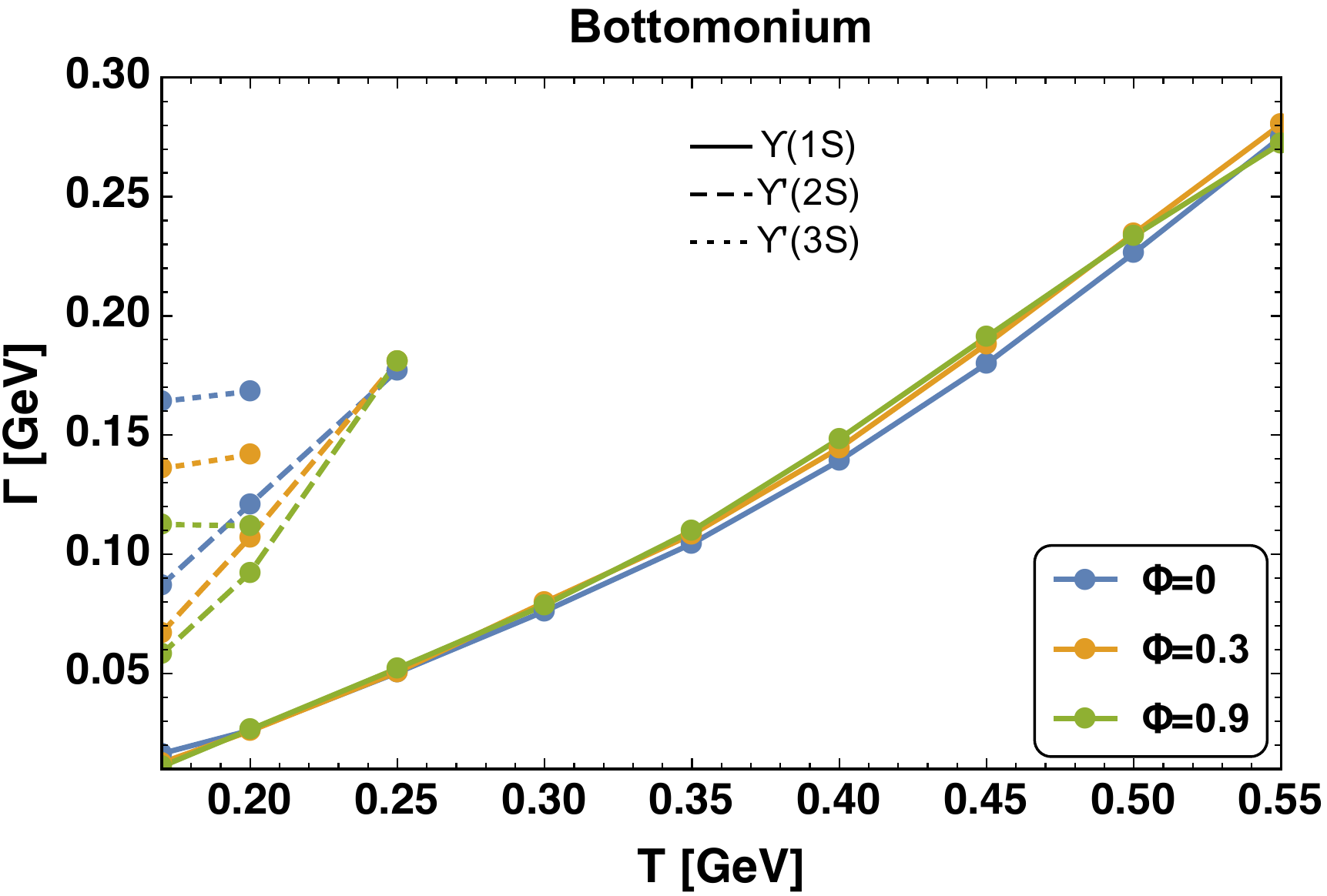}
		}
		\subfigure{
			\includegraphics[width=7.8cm]{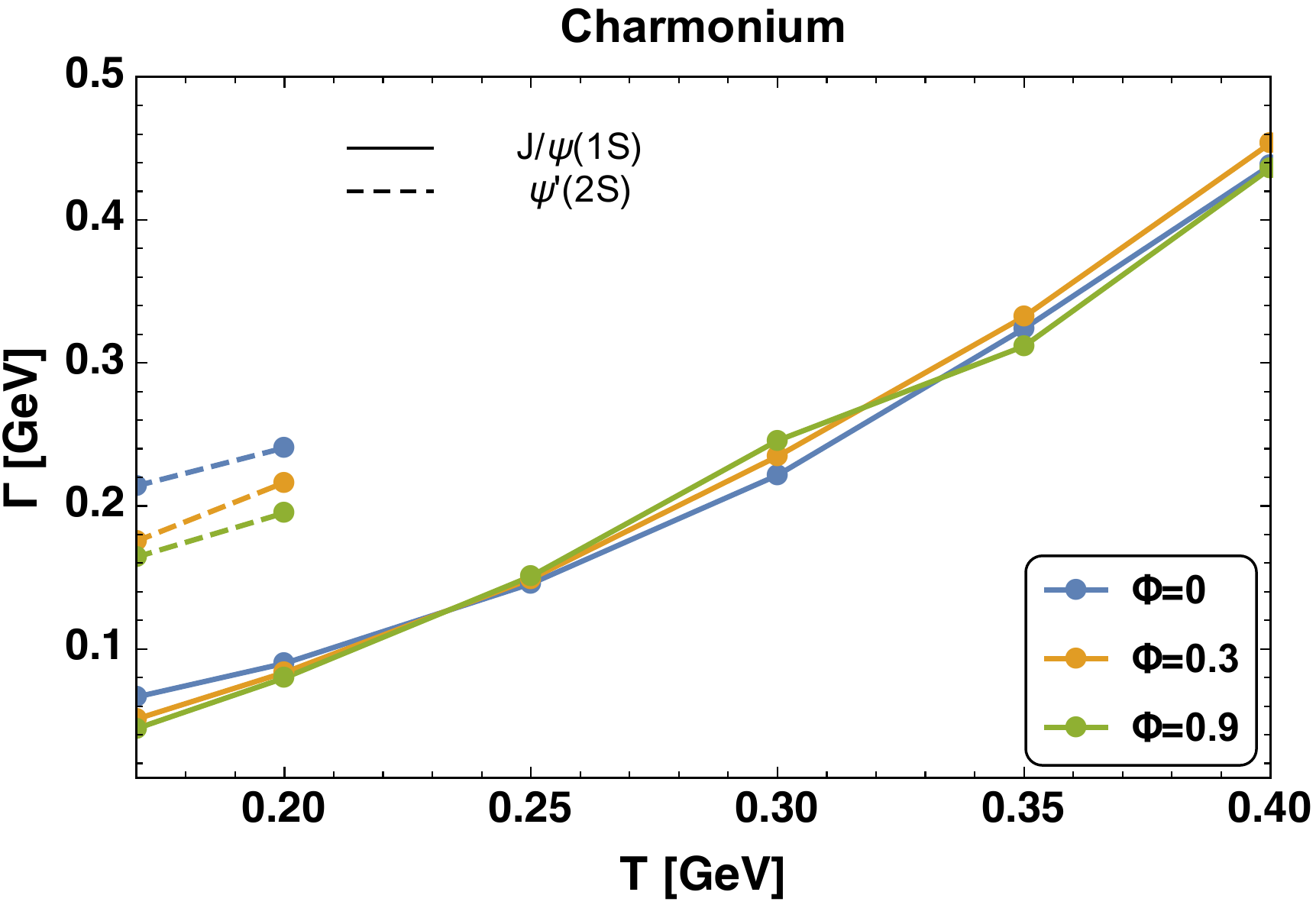}
		}
		\caption{
			Decay widths of the ground and excited states of bottomonium (left) and charmonium (right) as a function of temperature for different values of $ \Phi $.
		}
		\label{DecayW}
	\end{figure}

	In contrast, 
	the behaviour of the decay widths is more complicated. 
	In figure~\ref{DecayW}, 
	we plot the decay widths 
	of the ground and excited states of bottomonium and charmonium
	for different values of $\Phi$. 
	For the excited states ($\Upsilon^{\prime}(2S)$,$\Upsilon^{\prime}(3S)$, $\psi^{\prime}(2S)$), 
	the decay widths are decreasing functions of $\Phi$, 
	which is in contrast to our previous  computations~\cite{Thakur:2020ifi}. 
	However, it does not show a clear tendency for the ground states ($\Upsilon (1S)$,~$J/\psi(1S)$) 
	of both bottomonium and charmonium. 
	This behaviour of decay widths can be explained from
	the following two competing effects: 
	\begin{itemize}
		\item[1]
		The wave function becomes more spread 
		for larger values of $\Phi$. 
		When $|{\rm Im}\,V|$ is small compared to the real part, 
		the decay width $\Gamma$ is related to 
		the wave function $\psi$ by 
		$\Gamma \sim - \int  \psi^{\ast}({\rm Im} V)\psi \, d{\bf x}$, 
		if $\Phi$ does not have much effects on ${\rm Im} V$,  then the spreading of the wave function 
		increases the decay width. 
		\item[2] 
		Deformation of ${\rm Im }V$ by $\Phi$. 
		From figure~\ref{fig:potT350}, 
		we can see that the magnitude of ${\rm Im }V$ 
		is suppressed at finite $\Phi$ for large $r$. 
	\end{itemize}
	The excited states are larger in size
	compared to the ground states, 
	and are more sensitive to the larger-$r$ part of the $|{\rm Im }V|$,  
	and the effect 2 is more important. 
	As a result, the decay widths of the excited states 
	are decreasing functions of $\Phi$. 
	On the other hand, 
	the wave function of the ground states are more compact, 
	and 
	the imaginary part of the potential is not much affected by $\Phi$ in this region of $r$. 
	Hence, none of the two effects is dominant in this case.

	This behaviour differs from our previous results~\cite{Thakur:2020ifi}, 
	where the decay widths were increasing functions of $\Phi$
	for both ground and excited states. 
	The difference arises because of the different behaviour of ${\rm Im} V$ with $\Phi$ at small and large $r$. 
	In ref.~\cite{Thakur:2020ifi}, 
	when we have a bulk viscous correction $\Phi > 0$, 
	$|{\rm Im} V|$ was 
	enhanced    at small $r$ and 
	suppressed  at large $r$. 
	The origin of this difference is 
	attributed to the existence of a quasiparticle mass, 
	which was absent in ref.~\cite{Thakur:2020ifi}.

	Thus, we found that, 
	if we look at the decay widths, the excited states 
	are more sensitive to the bulk viscous correction, compared to the ground states. 
	This feature is potentially useful for the search of the critical point, at which the bulk viscous effect is going to be enhanced. 
	It would be interesting to 
	compare the collision energy dependence of $R_{AA}$ 
	of the ground and excited states.

	%%%%%%%%%%%%%%%%%%%%Fig%%%%%%%%%%%%%%%
	%%%%%%%%%%%%%%%%%%%%%%%%%%55
	
\begin{figure}[tb]
		\begin{center}
			%	\subfigure{
			\hspace{-8mm}
			\includegraphics[width=7.8cm]{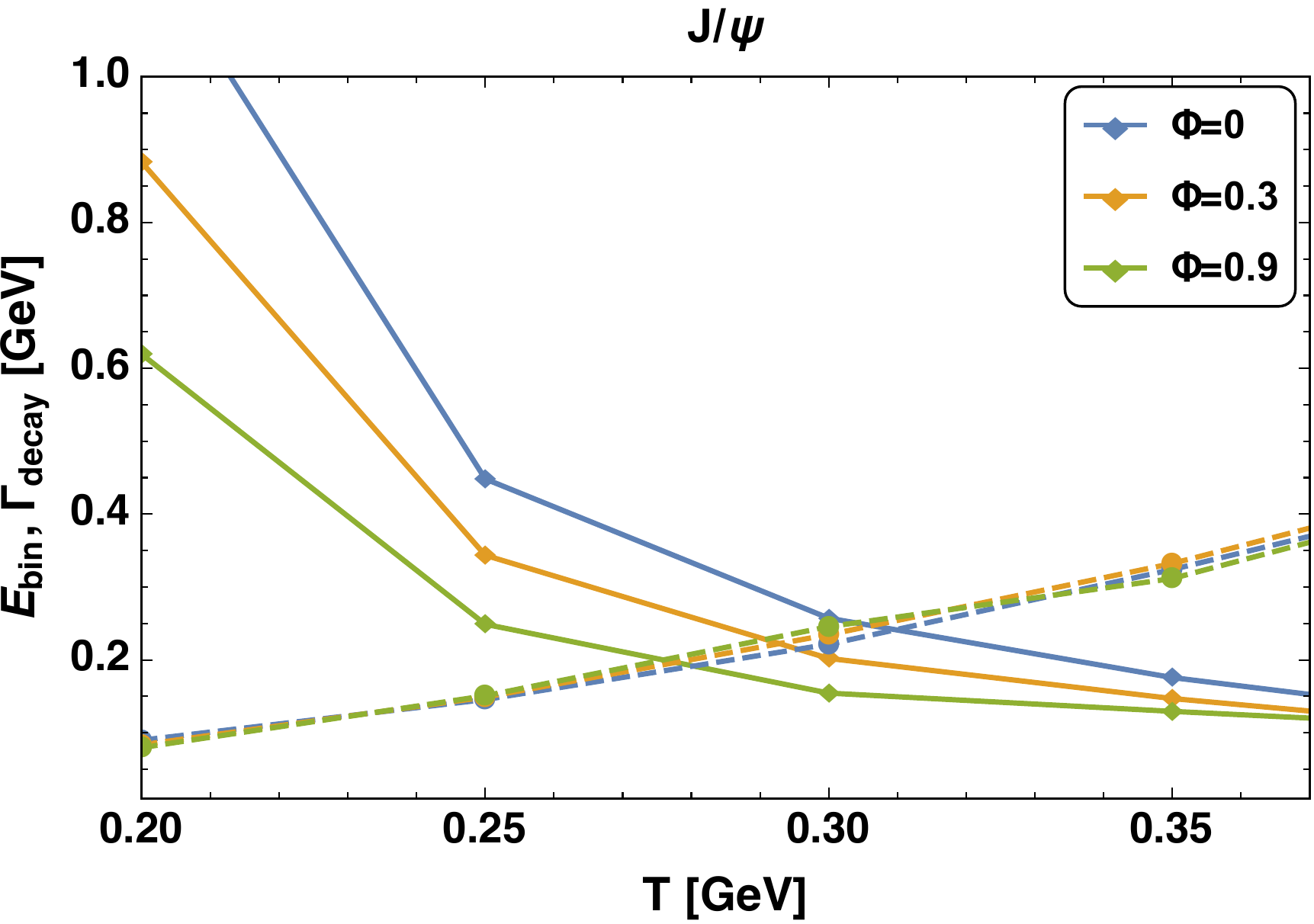}
			%	}
			%	\subfigure{
			%		\includegraphics[width=7.8cm]{figSF/EbindecaywidthJpsi}
			%	}
			\caption{
				The binding energies and decay widths of $J/\psi$ as a function of $ T $ for different values of $ \Phi $.
			}
			\label{EbinDWjpsi}
		\end{center}
	\end{figure}

	In figure~\ref{EbinDWjpsi}, we show the 
	binding energies and decay widths of the $ J/\psi $ as a function of $T$ for different values of $\Phi$. 
	From the intersection of $ E_{\rm bin} $ and $ \Gamma $, 
	we can read off the melting temperatures of the quarkonium states. 
	For a finite $\Phi$, the melting temperature goes down, 
	which is mainly driven by the decrease in the binding energy. 
	
	%%%%%%%%%%%%%%%%%%%%%%%%%%%%%%%%

	\section{Phenomenological implications}\label{sec:implications}
	
	In the previous section, we have discussed how the physical properties of heavy quarkonia are affected by the bulk viscous effect. 
	To discuss their experimental implications, 
	we need to relate them to the observables in heavy ion collisions. 
	In this section, we discuss the effects of bulk viscous correction on 
	physical observables, $ \psi^{\prime}/J/\psi$ ratio and $R_{AA}$.

	\subsection{Relative production yield of $ \psi^{\prime}$ to $J/\psi$} \label{subsec:production ratio}

	Here, we compute the ratio of the yield of 
	$\psi^{\prime}$ to $J/\psi$, following the method adopted in  refs.~\cite{Burnier:2015nsa,Lafferty:2019jpr}. 
	The dilepton production rate 
	is written by the spectral function as~\cite{McLerran:1984ay}
	\begin{equation}
		\frac{dR_{l\bar{l}}}{dK^4}=-\frac{Q_q \alpha_e^2}{3\pi^2K^2}n_B(k_0)\rho^{V}(K),
		\label{dileptonrate}
	\end{equation}
	where 
	$Q_q$ is the electric charge of the heavy quark in units of $ e $, 
	$ n_B $ is the Bose-Einstein distribution, 
	$ \alpha_e $ is the fine structure constant,
	and $ K= ({k^0, {\bf k}} )$ is the four momentum. 
	The decaying into dileptons occurs in the vacuum, 
	and we have to carry over the information of the medium 
	to the vacuum spectral function. 
	To achieve this, we perform an instantaneous freeze out at $ T=T_c $, 
	under which the in-medium spectral function is 
	projected to the vacuum ones.

	Let us walk through the procedure. 
	We start with the in-medium lepton-pair production rate~\eqref{dileptonrate}. 
	%
	%%%%%%%%%%%%%%%%%%%%%%%%%%%%%%%Figure%%%%%%%%%%
	\begin{figure}[tb]
		\begin{center}
			\subfigure{
				\hspace{-8mm}
				\includegraphics[width=7.8cm]{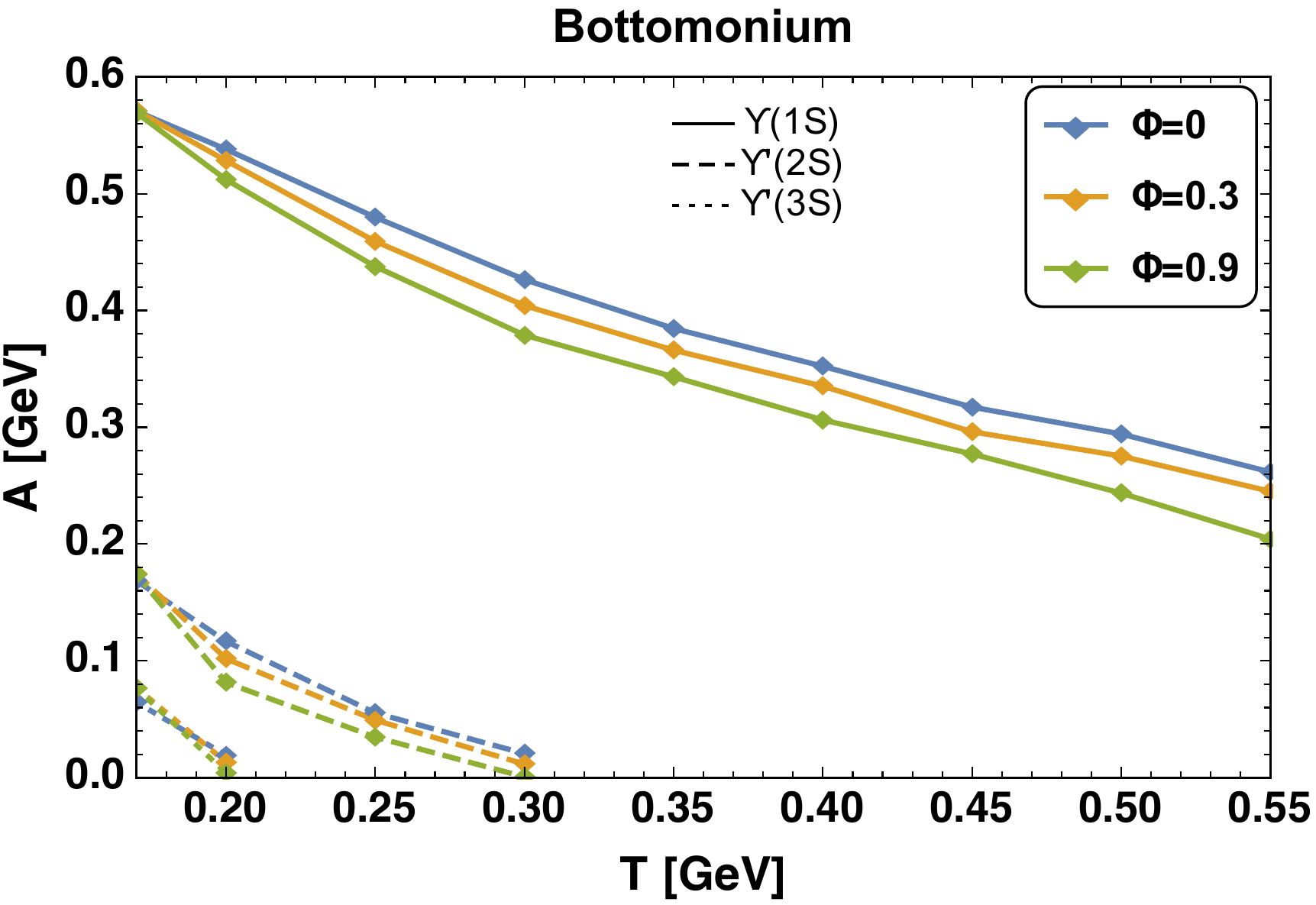}
				%	}
				%	\subfigure{
				\includegraphics[width=7.8cm]{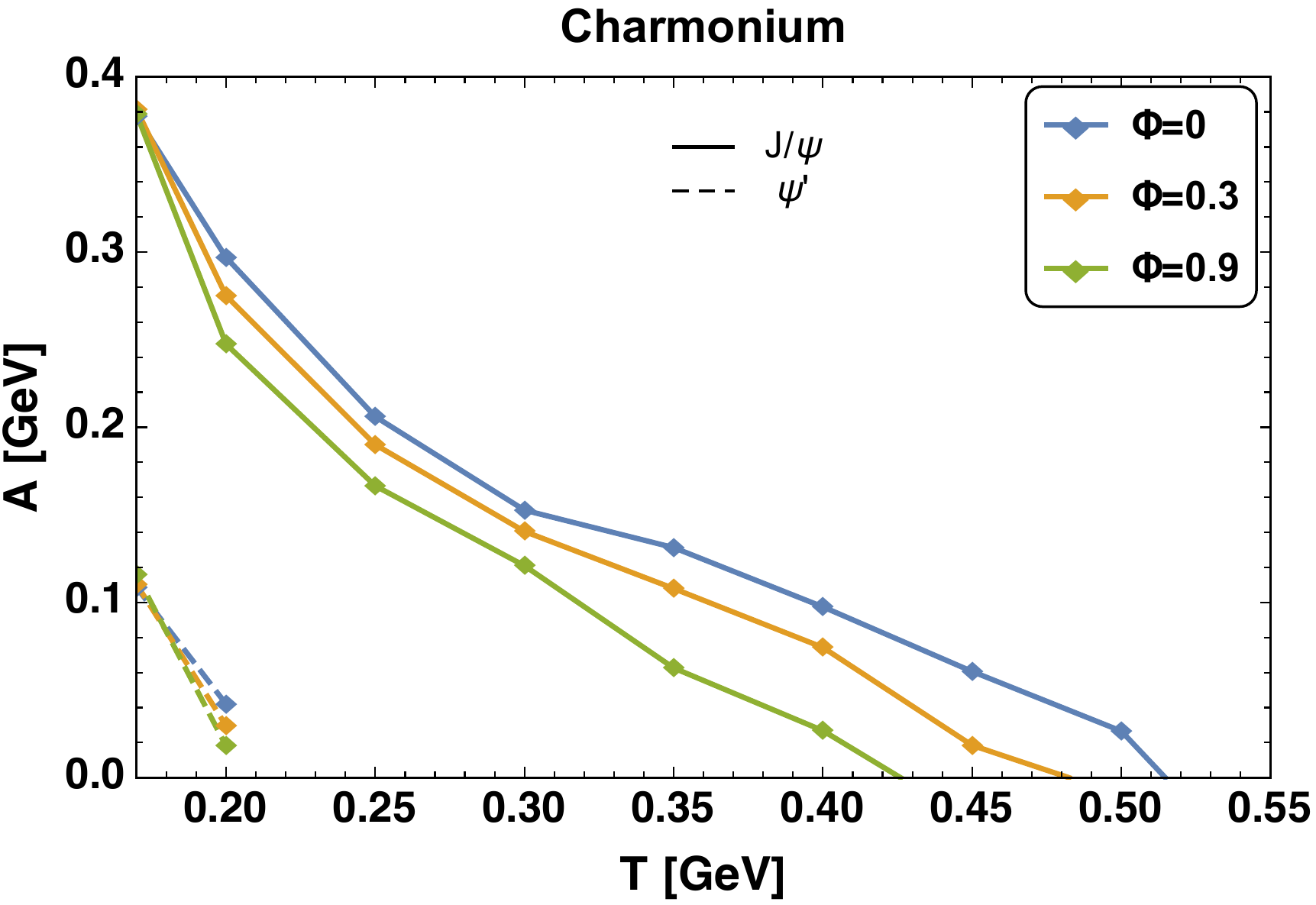}
			}
			\caption{
				Area under the bound-state peaks in the  bottomonium (left) and  charmonium (right) spectrum for different values of $ \Phi $.
			}
			\label{Intarea}
		\end{center}
	\end{figure}
	%%%%%%%%%%%%%%%%%%%%%%%%%%%%%%%%%%%%%%%%
	Since we are interested in the ratio of production yields, 
	prefactors can be dropped, and 
	the yield is proportional to 
	\begin{equation}
		R_{l\bar{l}}\propto \int dk_0 d^3{\bf k}~n_B(k_0)\frac{\rho^{V}(K)}{K^2}.
		\label{dilepton}
	\end{equation}
	After the change of variables from $K$ to $ \omega(=\sqrt{k_0^2-\bf{k}^2 }) $,
	the above equation becomes
	\begin{equation}
		R_{l\bar{l}}\propto \int d\omega d^3{\bf k}~n_B(\sqrt{\omega^2+\bf{k}^2})\frac{\rho^{V}(\omega)}{\omega^2}\frac{\omega}{\sqrt{\omega^2+\bf{k}^2}}. 
	\end{equation}
	Each peak in $ \rho^{V}(\omega)/\omega^2 $ 
	gives the contribution from each bound state. 
	We fitted each peak structure with the skewed Breit-Wigner (\ref{BW}), 
	and determined the resonance mass, $ M_n $, and the decay width. 
	We also computed the area $A_n$ under each spectral peak. 
	The projection process mentioned above is implemented by 
	replacing the in-medium peaks with delta-function peaks, 
	\begin{equation}
		\rho^{V}(\omega)/\omega^2=\sum_n A_n \delta(\omega-M_n) . 
	\end{equation}
	After performing the integration over $\omega$, 
	the contribution from the state $n$ is written as 
	\begin{equation}
		R_{l\bar{l}}\propto A_n \int  d^3{\bf k}~n_B(\sqrt{M_n^2+\bf{k}^2})\frac{M_n}{\sqrt{M_n^2+\bf{k}^2}}.
		\label{Rlleq}
	\end{equation}
	%%%%%%%%%%%%%%%%%%%%%%%%%%%%%
	\begin{figure}[tb]
		\begin{center}
			%\subfigure{
			%	\hspace{-8mm}
			\includegraphics[width=15cm]{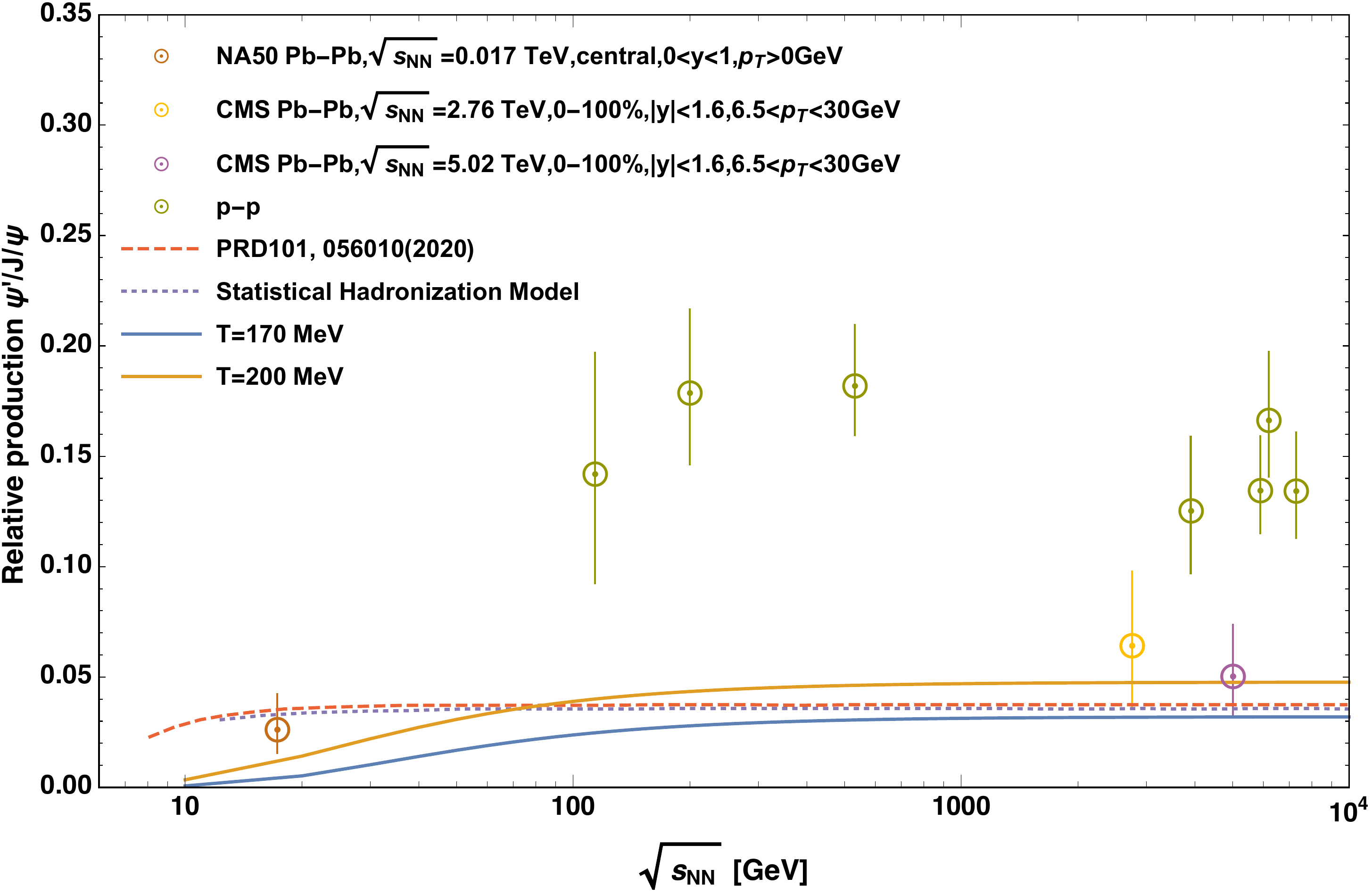}
			%	\includegraphics[width=7.8cm]{figSF/RatiopsibyjpsiT17}
			%	}
			%	\subfigure{
			%	\includegraphics[width=7.8cm]{figSF/RatiopsibyjpsiT2}
			%	}
			\caption{
				Relative production yield $ \psi^{\prime}/ J/\psi $ ratio as a function of $ \sqrt{s_{NN}} $, where two solid lines correspond to different temperatures
				at which the areas and masses are computed ($T=170$ MeV and $T=200$ MeV)(this work).
				%	
				%ratio at $ T =170$~MeV and $ T =200$~MeV .
				The red dashed line shows Gauss law potential model~\cite{Lafferty:2019jpr} and purple dotted line shows statistical hadronization model prediction~\cite{Andronic:2017pug}. The data points for  Pb-Pb collisions are from the experiments for NA50~\cite{NA50:2006yzz} (brown symbol) %, ALICE~\cite{ALICE:2015jrl} (blue symbol), 
				and CMS~\cite{CMS:2014vjg,CMS:2016wgo} Collaborations (yellow and purple symbols). The data points for p-p collisions are from experiments at SPS,  RHIC, and the LHC~\cite{Andronic:2017pug, Drees:2017zcb, NA51:1998uun, PHENIX:2016vmz,LHCb:2011zfl,LHCb:2012geo,ALICE:2017leg} (green symbol). %[117–122]for the pp baseline ~\cite{Drees:2017zcb} (green symbol).
			}
			\label{rate}
		\end{center}
	\end{figure}
	To get the total number density, 
	we also have to divide by the electromagnetic decay rate 
	of a vacuum state, which is proportional to 
	the squared wave function at the origin over the mass squared~\cite{Bodwin:1994jh}. 
	The final expression of the ratio is 
	\begin{equation}
		\frac{N_{\psi^{\prime}}}{N_{J/\psi}}
		=
		\frac{	R_{l\bar{l}}^{\psi^{\prime}}}{R_{l\bar{l}}^{J/\psi}}\cdot \frac{M_{\psi^{\prime}}^2|\psi_{J/\psi}(0)|^2}{M_{J/\psi}^2|\psi_{\psi^{\prime}}(0)|^2}.
		\label{eq:ratio}
	\end{equation}
	We here take $ M_{\psi^{\prime}}=3.684 $ GeV and $ M_{J/\psi}=3.0969 $ GeV, $ \psi_{J/\psi}(0)= 1.454 $ GeV$ ^3 $, 
	and $\psi_{\psi^{\prime}}(0)=0.927 $ GeV$^3$~\cite{eichten1995quarkonium}.
	Thus, in this procedure, 
	the ratio depends on 
	the resonance mass $M_n$ and the area under the peak $A_n$, 
	both of which are affected by medium. 
	It also depends on the temperature at which this ``freezing'' occurs.

	In figure~\ref{Intarea}, we plot the integrated areas 
	under the bound state peaks for bottomonium and charmonium states. A general trend is that the peak area 
	becomes smaller for larger $T$ and $\Phi$. 
	The proportionality change in the area of the excited state is more pronounced compared to the ground state. 
	Thus, the effect of bulk viscous correction is more significant on the peak areas of the excited states 
	compared to those of the ground states, which is similar to the case of in-medium masses and decay widths. 
	
	% At fixed high $ T $, this effect is more pronounced for the ground state which can be understood from the behaviour of spectral functions of bottomonium and charmonium states at fixed high T for different values of $ \Phi $ (figure \ref{SFT350phi}). From figure~\ref{SFT350phi}, it is clear that the effect of  $ \Phi $ is more significant on the ground states, whereas the excited states are already dissolved at such a high temperature. The spectral peaks get broadened and shifted towards lower values of frequency $ \omega $ from their central values, which results in the decrease in  the  peak areas  with increase in $ \Phi $. Hence the effect of $ \Phi $ is more pronounced on the peak areas of the ground states as compared to the excited states at fixed high $ T $.

	In order to connect the ratio with 
	the collision energies, we use the freeze-out temperature 
	(which is used in the Bose distribution in eq.~\eqref{Rlleq}) fitted to reproduce the particle yields~\cite{Andronic:2017pug}, 
	\begin{equation}
		T(\sqrt{s_{NN}})=
		\frac{158 {\rm MeV}}
		{1+{\rm exp}(2.60-{\rm ln}(\sqrt{s_{NN}})/0.45)}, 
	\end{equation}
	where $ \sqrt{s_{NN}} $ is 
	the numerical value of the center-of-mass energy in GeV. 
	We here set the baryon chemical potential to zero. 
	After substituting all the values in eq.~(\ref{eq:ratio}), we can compute the  $ \psi^{\prime}/ J/\psi$ ratio over a range of center-of-mass energies.
	%  %%%%%%%%%%%%%%%%%%%%%%%%%%%%%
	% \begin{figure}[tb]
	% 	\begin{center}
	% 		%\subfigure{
	% 		%	\hspace{-8mm}
	% 		\includegraphics[width=10cm]{figSF/RatiowithphidiffT.pdf}
	% 		\caption{Variation of relative production yield $ \psi^{\prime} / J/\psi $ ratio with the bulk viscous correction $\Phi$.}
	% 		\label{ratio}
	% 	\end{center}
	% \end{figure}
	% %%%%%%%%%%%%%%%%%%%%%%%%%%%%%%%%%%%%%%%%%%%
	%%%%%%%%%%%%%%%%%%%%%%%%%%%%%
	\begin{figure}[tb]
		\begin{center}
			%\subfigure{
			%	\hspace{-8mm}
			\includegraphics[width=10cm]{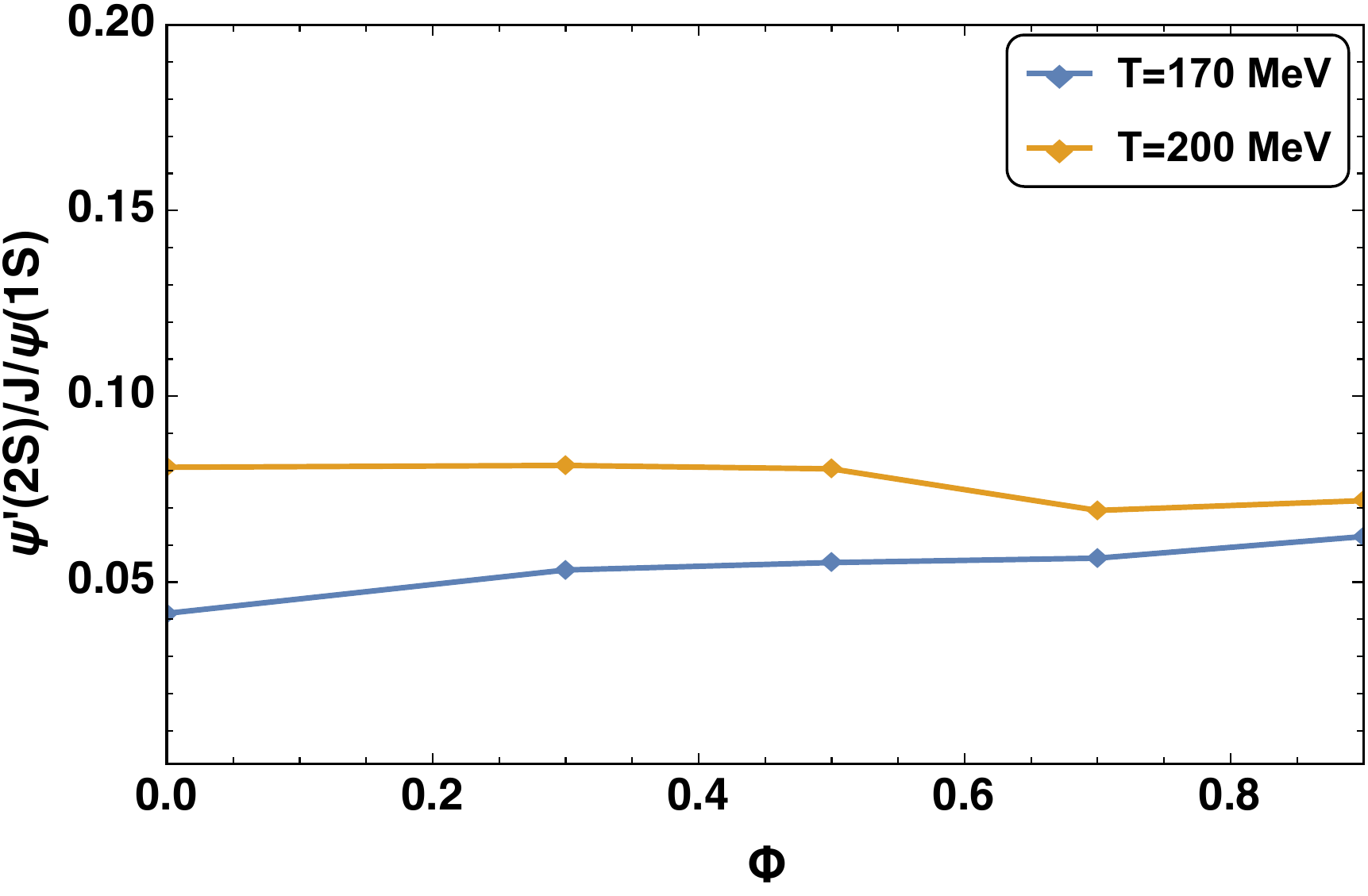}
			\caption{Relative production yield $ \psi^{\prime} / J/\psi $ ratio as a function of $\Phi$.}
			\label{ratio}
		\end{center}
	\end{figure}
	%%%%%%%%%%%%%%%%%%%%%%%%%%%%%%%%%%%%%%%%%%%
	Note that our approach applies for the direct production of $J/\psi$, and the feed down is not considered. 
	
	Figure~\ref{rate} shows the 
	relative production yield
	$ \psi^{\prime}/J/\psi$ ratio as a function of center-of-mass energy $ \sqrt{s_{NN}} $. 
	We compare our results with 
	the statistical hadronization model~\cite{Andronic:2017pug}, 
	the Gauss law potential model~\cite{Lafferty:2019jpr}, 
	and experimental data for Pb-Pb~\cite{NA50:2006yzz,ALICE:2015jrl,CMS:2014vjg,CMS:2016wgo} 
	and $pp$~\cite{Andronic:2017pug,Drees:2017zcb,NA51:1998uun,PHENIX:2016vmz,LHCb:2011zfl,LHCb:2012geo,ALICE:2017leg} collisions. 
	For both of the freezing temperatures, $170$ MeV and $200$ MeV, our results are close to those 
	from the statistical hadronization model and
	the Gauss law potential model above 100 GeV center-of-mass energy. % $ \sqrt{s_{NN}} $

	In figure~\ref{ratio}, we plot the 
	$ \psi^{\prime}/ J/\psi$ ratio as a function of $\Phi$. 
	The two lines correspond to different freezing temperatures
	at which the spectral function is projected 
	to the vacuum one 
	($T=170$ MeV and $T=200$ MeV). %
	We find that the ratio moderately increases as a function of $\Phi$ at $T=170$ MeV, 
	whereas it does not show a clear tendency at $T=200$ MeV. 
	The behaviour of $ \psi^{\prime}/J/\psi$ ratio as a function $\Phi$ can be understood from the behaviour of $R_{l\bar{l}}$   
	as a function of $\Phi$ for both $J/\psi$ and $\psi^{\prime}$. 
	From eq.~(\ref{Rlleq}), we can see that $R_{l\bar{l}}$ depends on two factors, i.e., areas $ (A)$ and masses $ (M)$. 
	The areas and masses decrease as a function of $\Phi$ for both the $J/\psi$ and $\psi^{\prime}$. 
	The decrease in areas 
	as a function of $\Phi$ results in the decrease in
	$R_{l\bar{l}}$
	with $\Phi$.  Whereas, the decrease in masses as a function of $\Phi$  results in the increase in $R_{l\bar{l}}$
	with $\Phi$.
	At $T=170$ MeV, the areas are almost the same (figure~\ref{Intarea}) 
	and 
	the effects of $\Phi$ on $R_{l\bar{l}}$ arises only because of the shift in the masses as a function of $\Phi$. 
	The decrease in the masses of $ \psi^{\prime}$ as a function of $\Phi$ is sharper as compared to %the shift in the masses of 
	$J/\psi$, hence $R^{\psi^{\prime}}_{l\bar{l}}/R^{J/\psi}_{l\bar{l}}$ ratio (eq.~\eqref{eq:ratio}) increases which results in the increase in $ \psi^{\prime}/ J/\psi$ ratio as a function of $\Phi$. 
	However, at $T=200 $ MeV, 
	the effects of $\Phi$ on $R^{\psi^{\prime}}_{l\bar{l}}/R^{J/\psi}_{l\bar{l}}$ ratio arises due to both the shift in areas as well as the shift in the masses as a function of $\Phi$.
	Hence, $ \psi^{\prime}/ J/\psi$ ratio is not showing monotonous behaviour as a function of $\Phi$.

	\subsection{Nuclear modification factor $R_{AA}$}\label{subsec:RAA}
	
	In this subsection, we give a rough estimate of the nuclear modification factor $R_{AA}$ to see how it gets 
	affected by the bulk viscous correction.
	The analysis here closely follows ref.~\cite{Blaizot:2021xqa}. 
	To compute the $R_{AA}$, 
	we first compute the survival probability of a heavy quarkonium state.  
	We assume that the longitudinal dynamics of the plasma is the Bjorken expansion, and ignore the transverse expansion. 
	We also assume that a produced quarkonium does not move in the transverse plane. 
	We denote the probability that a heavy quarkonium state 
	is in state $n$ by $p_n$. 
	We assume that initially the heavy quarkonium is in state $n$, 
	$p_n(\tau_0)=1$, where 
	$\tau_0$ is the proper time at which the system 
	start interacting with the medium. 
	We treat the interaction of quarkonia with the medium 
	in an adiabatic approximation, on the assumption 
	that the medium evolves slowly. 
	Then, 
	the fate of the state is determined by the 
	decay rate $\Gamma(T(\tau))$ 
	determined by the local temperature at proper time $\tau$. 
	Under those assumptions, 
	the time evolution of $p_n(\tau)$ is given by 
	\begin{equation}
		\frac{{\rm d}}{{\rm d}\tau}p_n (\tau) = -\Gamma(T(\tau)) p_n(\tau). 
	\end{equation}
	We choose the initial time as $\tau_0=0.6 ~{\rm fm}$
	and 
	the initial condition is $p_n(\tau_0) = 1$. 
	The survival probability of a given quarkonium state is $S=p_n(\tau_f)$, where $\tau_f$ is the final interaction time, 
	which is determined as the time at which 
	the temperature reaches a certain value, $T = T_f$. 
	For an analytical tractability, 
	we fit the decay width 
	(computed in subsection~\ref{sec:spectral function}) 
	with the following function, 
	\begin{equation}
		\Gamma = aT+bT^{2},
		\label{gmmafit}
	\end{equation}
	where $a$ and $b$ are parameters. 
	The survival probability depends on 
	the position in the transverse plane and the impact parameter, 
	$S = S(\bf s, \bf b)$, 
	because of the initial temperature profile. 
	We choose the origin of coordinates
	in the transverse plane to coincide with the center of one nucleus.
	The initial energy density, $ \epsilon_0 $, 
	in the transverse plane is assumed to be proportional to the density of the participants 
	and can be computed by using the optical Glauber model~\cite{Blaizot:1988ec,Kolb:2000sd,Miller:2007ri},  
	\begin{equation}
		\epsilon_0({\bf b},{\bf s})=
		K\left\{T_A({\bf s})
		\left[
		1-\left(1-\frac{\sigma T_A({\bf{s-b}})}{A}\right)^{A}
		\right]+
		T_A({\bf s-b})
		\left[
		1-\left(1-\frac{\sigma T_A({\bf{s}})}{A}\right)^{A}
		\right]\right\},
	\end{equation}
	where $K$ is a proportionality coefficient.
	From the relation between the energy density and the temperature, $ T \sim \epsilon^{1/4}$, 
	one can obtain the  initial temperature profile $ T_{0}(\bf b,\bf s) $ for a given impact parameter $\bf b$ as 
	\begin{equation}
		T_0({\bf b},{\bf s})=
		T_0(\mathbf 0,\mathbf 0)
		\left(
		\frac{T_A({\bf s})
			\left[
			1-\left(1-\frac{\sigma T_A({\bf{s-b}})}{A}\right)^{A} \right]+T_A({\bf s-b})
			\left[
			1-\left(1-\frac{\sigma T_A({\bf{s}})}{A}\right)^{A}
			\right]
		}{
			T_A(\mathbf 0,\mathbf 0 )
			\left[
			1-\left(1-\frac{\sigma T_A(\mathbf 0)}{A}\right)^{A}
			\right]+T_A(\mathbf 0,\mathbf 0 )
			\left[
			1-\left(1-\frac{\sigma T_A(\mathbf 0)}{A}\right)^{A}
			\right]
		}\right)^{1/4}, 
	\end{equation}
	%
	% %%%%%%%%%%%%%%%%%%%%%%%%%%%%%%%%%%%%%%%%%%%%%%%
	%%%%%%%%%%%%%%%%%%%%%%%%%%%%Figure%%%%%%%%%%%%%%%%%%%%%%
	\begin{figure}[tb]
		\begin{center}
			\subfigure{
				\hspace{-8mm}
				\includegraphics[width=7.8cm]{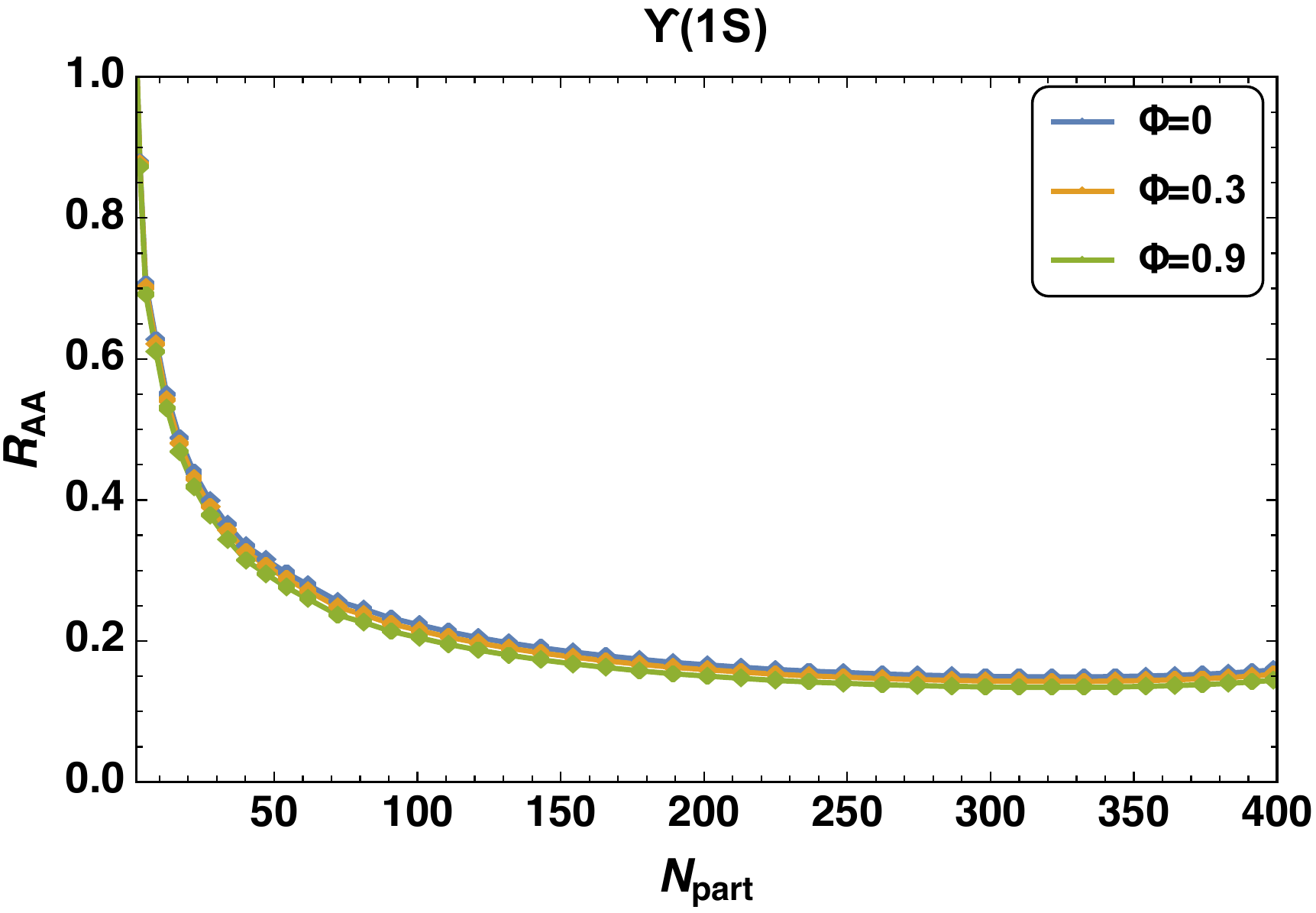}
				%	}
				%	\subfigure{
				\includegraphics[width=7.8cm]{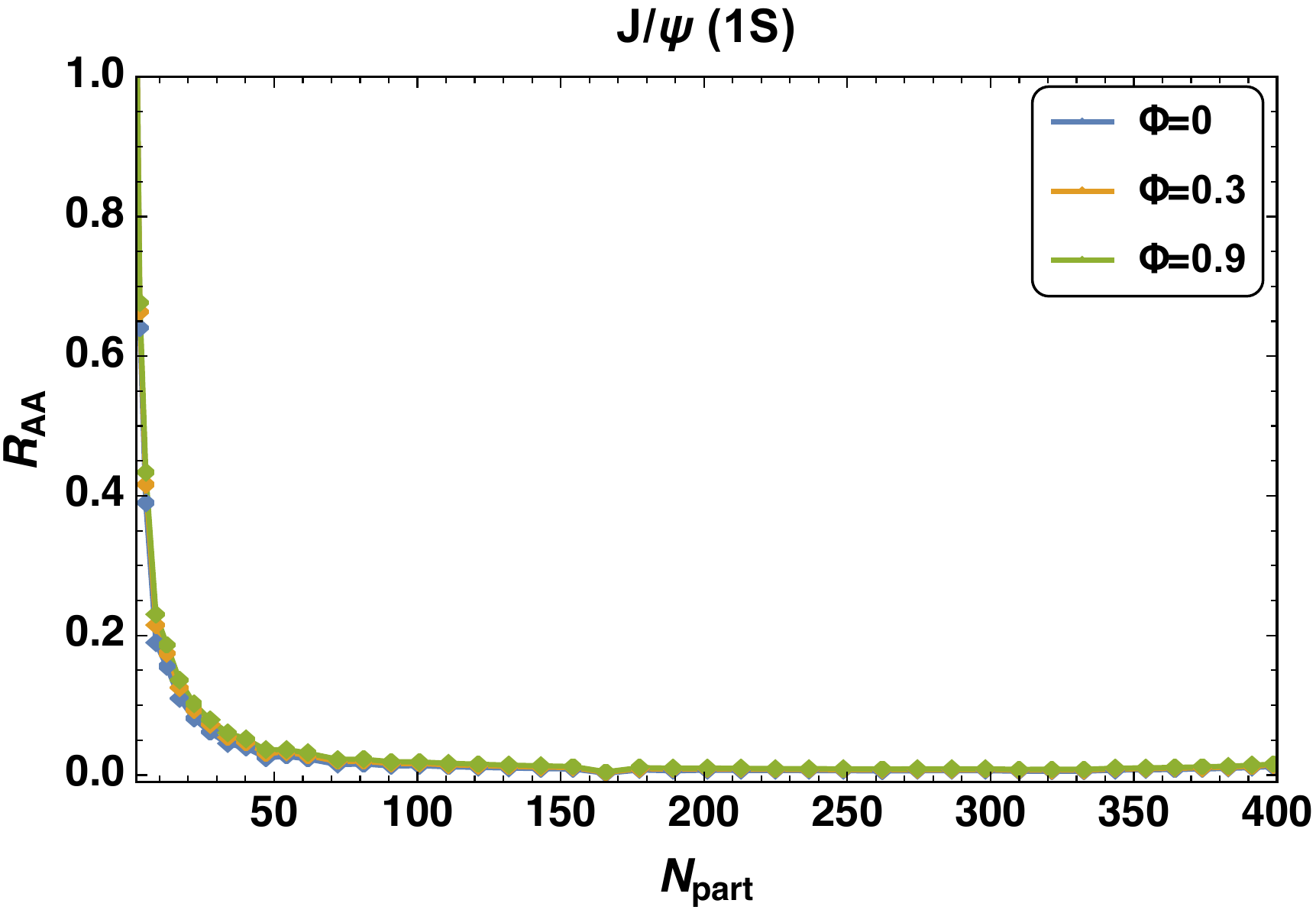}
			}
			\caption{
				Nuclear modification factor $R_{AA}$ %($r_{AA}({\rm PbPb})/r_{AA}({\rm pp})$) 
				as a function of number of participants $N_{\rm part}$ for $\Upsilon (1S)$(left) and $J/\psi$ (right) for different values of $ \Phi $ for $T_f=158$ MeV.
			}
			\label{RAAfig}
		\end{center}
	\end{figure}
	%%%%%%%%%%%%%%%%%%%%%%%%%%%%Figure%%%%%%%%%%%%%%%%%%%%%%
	%
	where $T_A$ is the nuclear thickness function and $T_0(\mathbf 0,\mathbf 0)$ is the temperature in the center of the overlapping region for the most central collisions. 
	We consider collisions at the LHC
	with $\sqrt{s}=5.02$~TeV, 
	and take $T_0(\mathbf 0,\mathbf 0)=500 $~MeV~\cite{Alberico:2013bza,Alqahtani:2020paa},
	$A=207$, $\sigma=70 $~mb \cite{Loizides:2014vua}. 
	We use the thickness function of a nucleus 
	with uniform density inside a sphere, 
	\begin{equation}
		T_A({\bf s})=\frac{3}{2\pi r_0^3}(R_A^2-{\bf s}^2)^{1/2},
	\end{equation}
	where $R_A=r_0A^{1/3}$ and $r_0=1.2$ fm. 
	The time dependence of the temperature is given by 
	\begin{equation}
		T(\tau,{\bf b},{\bf s}) = T_0({\bf b},{\bf s})\left(\frac{\tau_0}{\tau}\right)^{1/3}.
	\end{equation}
	Using the parametrization of the decay width (\ref{gmmafit}), 
	the survival probability can be computed analytically as 
	\begin{equation}
		S({\bf b},{\bf s})=e^{-1.5 a T_0({\bf b},{\bf s})\tau_0\left(\left(\frac{T_0({\bf b},{\bf s})}{T_f}\right)^2-1\right)-3bT_0({\bf b},{\bf s})^2 \tau_0\left(\frac{T_0({\bf b},{\bf s})}{T_f}-1\right)}.
	\end{equation}
	Here, we choose the final temperature to be $T_f=158$~MeV. 
	For a given impact parameter, we compute the following quantity, 
	\begin{equation}
		r({\bf b})=\frac{\int d^2{\bf s}T_A({\bf s})T_A({\bf s-b})S({\bf b},{\bf s})}{\int d^2{\bf s}T_A({\bf s})T_A({\bf s-b})}.
	\end{equation}
	The nuclear modification factor $R_{AA}$ for a given impact parameter 
	is computed as
	\begin{equation}
		R_{AA}=\frac{r (\bf b)}{r({\bf b} = {\bf b}^\ast)}, 
	\end{equation}
	where ${\bf b}^\ast$ is the impact parameter at which 
	the number of participants, $N_{\rm part}$, is two. 
	We have introduced this normalization 
	so that $R_{AA}$ approaches unity when $N_{\rm part}=2$. 
	We compute the nuclear modification factor $R_{AA}$ 
	as a function of $N_{\rm part}$~\cite{Blaizot:1988hh}, 
	which can be written as 
	% \begin{equation}
	%     N_{\rm part}({\bf b})
	%     =\int_{S_{\rm eff}}d^2{\bf s}\left[T_A({\bf s})+T_A({\bf s-b})\right],
	% \end{equation}
	% or
	\begin{equation}
		N_{\rm part}({\bf b})=
		\int d^2{\bf s}\left\{T_A({\bf s})
		\left[
		1-\left(1-\frac{\sigma T_A({\bf{s-b}})}{A}\right)^{A}
		\right]+
		T_A({\bf s-b})
		\left[
		1-\left(1-\frac{\sigma T_A({\bf{s}})}{A}\right)^{A}
		\right]\right\},
	\end{equation}
	% where $S_{\rm eff}$ is the area in the transverse plane over which the two nuclei overlap.
	%
	Figure~\ref{RAAfig} shows the computed 
	nuclear modification factor $R_{AA}$ 
	as a function of number of participants $N_{\rm part}$ for $\Upsilon$ and $J/\psi$. 
	Different lines correspond to different values of $ \Phi $. 
	$R_{AA}$ shows a monotonous decrease as a function of $N_{\rm part}$ for both states, as expected. 
	$R_{AA}$ decreases slightly as a function $\Phi$ for $\Upsilon$ and increase slightly for $J/\psi$. 
	We observe that $R_{AA}$s of the ground states are 
	not much affected by the bulk viscous correction $\Phi$. 
	Since $R_{AA}$ depends on the decay widths and decay widths of the ground states of quarkonia are not much affected by
	$\Phi$, as shown in figure~\ref{DecayW}.

	However, the decay widths of 
	excited states have a stronger dependence on the 
	bulk viscous correction, 
	as can be seen in figure~\ref{DecayW}. 
	Thus, we expect that the $R_{AA}$s of the excited states 
	are going to depend on $\Phi$ more strongly 
	than those of the ground states. 
	Within the current computational procedure, 
	we cannot make a reliable estimate of them, 
	since we cannot extrapolate the decay widths of the excited 
	states 
	to the temperatures ($\sim 500$ MeV) realized in the fluid. 
	In order to estimate $R_{AA}$s of the excited states, 
	a different method is needed. 
	We argue that it will be interesting to compare 
	the collision energy dependence 
	of the $R_{AA}$s of excited states 
	and ground states to look for the sign 
	of an enhancement of the bulk viscosity near the critical point.

	\section{Summary and discussions}\label{sec:summary} 
	
	In this work, we studied the effect of bulk viscous nature of a plasma on the properties of heavy quarkonia and 
	its implications on the experimental observables
	in heavy ion collisions. 
	We incorporated the bulk viscous correction, 
	whose strength is quantified by a parameter $\Phi$, 
	by deforming the distribution functions of thermal quarks and gluons. 
	We also introduced quasiparticle masses, 
	with which the behaviour of thermodynamic properties can be well reproduced near the crossover temperature. 
	We computed the dielectric permittivity of 
	a bulk viscous medium within the hard thermal loop approximation at one-loop, 
	and we have obtained an in-medium potential using the dielectric permittivity. 
	From the in-medium potential, 
	we computed the spectral functions, which encode physical properties of quarkonium states.

	To quantify the effects of a bulk viscous medium, 
	we have fitted each resonance peak of spectral functions 
	with a skewed Breit-Wigner-type function. 
	By doing this, we can extract quantities such as 
	in-medium masses and decay widths. 
	We found that the in-medium masses and binding energies of the quarkonium states decreases with the increase in bulk viscous correction. 
	The decay widths of the excited states decrease 
	as a function of $\Phi$. 
	On the other hand, 
	the decay widths of ground state do not show a particular tendency 
	as a function of $\Phi$. 
	This can be understood as a result of two competing effects, 
	namely 
	the widening of wave functions because of an increased screening, 
	which tends to increase the decay width, 
	and 
	the deformation of the imaginary part of the potential, 
	which tends to decrease the decay width. 
	For excited states, the latter one is dominant, 
	while for ground states the two effects are comparable.

	We also studied the implications of the bulk viscous effects on the physical observables 
	such as $ \psi' $ to $ J/\psi $ ratio 
	and the nuclear modification factor $ R_{AA}$. 
	The $ \psi^{\prime}/ J/\psi $ ratio 
	turned out to show a complicated dependence 
	on the bulk viscous parameter $\Phi$. 
	This appears as a result 
	of the combined effect of 
	the behaviour of the peak area and the resonance mass
	as a function of $\Phi$. 
	We also gave a rough estimate of 
	the nuclear modification factor $R_{AA}$ for $\Upsilon $ 
	and $J/\psi $ states. 
	We found that $R_{AA}$s of ground states 
	are not much affected by the bulk viscous correction $\Phi$. 
	However, we expect that the $R_{AA}$ of excited states 
	are more sensitive to $\Phi$, 
	since their decay widths have stronger dependence on $\Phi$ 
	than the ground states.
	Experimentally, 
	it will be interesting to see the collision energy dependence 
	of the $R_{AA}$s of excited states and ground states. 
	If there is an enhancement of the bulk viscosity near the critical point, that might result in a non-monotonic behaviour 
	of $R_{AA}$ of the excited states.

	%%%%%%%%%%%%%%%%%%%%%%%%%%%%%%%%%%%%%%%%%
	%%%%%%%%%%%%%%%%%%%%%%%%%%%%%%%%%%%%%%%%%
	\section{Acknowledgement}
	We thank Hyungjoo Kim for the discussions at the early stage of this work. Y.~H. and L.~T. are supported by the Korean Ministry of Education, Science and Technology, Gyeongsangbuk-do and Pohang City  
	at the Asia Pacific Center for Theoretical Physics (APCTP).
	L.~T. is supported by National Research Foundation (NRF) funded by the Ministry of Science of Korea (Grant No. 2021R1F1A1061387).
	Y.~H. is supported by the National Research Foundation (NRF) funded by the Ministry of Science of Korea (Grant No. 2020R1F1A1076267).
	%%%%%%%%%%%%%%%%%%%%%%%%%%%%%%%%%%%%%%%%%%%%%%%%%%%%%%%%%

%\bibliographystyle{utphys.bst}

%\bibliography{refs}
\providecommand{\href}[2]{#2}\begingroup\raggedright\endgroup

%
%
% 5 pages, 4 figures, 1 table, contribution to  HADRON 2021-19th International Conference on Hadron Spectroscopy and Structure in memoriam Simon Eidelman
\end{document}